\documentclass[smallcondensed]{svjour3}     % onecolumn (ditto)
\smartqed  % flush right qed marks, e.g. at end of proof
\usepackage{graphicx}
\usepackage{mathptmx}      % use Times fonts if available on your TeX system
\usepackage{amsmath}
\usepackage{latexsym}
\usepackage{physics}
\newcommand*{\blb}{{\big<}}
\newcommand*{\blk}{{\big>}}
\newcommand*{\bl}{{\big |}}
\journalname{FBS}

\begin{document}

\title{Theory of surrogate 
nuclear and atomic 
reactions 
%in a few-body approach 
with three charged particles in the final state
proceeding through a resonance in the intermediate subsystem  
%\thanks{Grants or other notes
%about the article that should go on the front page should be
%placed here. General acknowledgments should be placed at the end of the article.}
}

\titlerunning{Theory of surrogate nuclear and atomic reactions}        % if too long for running head

\author{A. M. Mukhamedzhanov        \and    A. S. Kadyrov}

\institute{A. M. Mukhamedzhanov \at
             Cyclotron Institute, Texas A\&M University, College Station, TX 77843 \\
              \email{akram@comp.tamu.edu}           
           \and
           A. S. Kadyrov \at
              Curtin Institute for Computation and Department of Physics and Astronomy, Curtin University, \\
              GPO Box U1987, Perth, WA 6845, Australia \\
\email{a.kadyrov@curtin.edu.au}
}

\date{Received: date / Accepted: date}
% The correct dates will be entered by the editor

\maketitle

\begin{abstract}
Within a few-body formalism, we develop a general theory of surrogate nuclear and atomic reactions with the excitation of a resonance in the intermediate binary subsystem leading to three charged particles in the final state. The Coulomb interactions between the spectator and the resonance in the intermediate state and between the three particles  in the final state are taken into account. Final-state 
three-body Coulomb multiple-scattering effects are accounted for using the formalism of the three-body Coulomb asymptotic states based on the work published by one of us (A.M.M.) under the guidance of L.~D.~Faddeev. An expression is derived for the triply differential cross section. It  can be used for investigation of the Coulomb effects on the resonance line shape as well as the energy dependence of the cross section. 
We find that simultaneous inclusion of the Coulomb effects in the intermediate and final state decreases the effect of the final-state Coulomb interactions on the triply differential cross section.
% \subclass{MSC code1 \and MSC code2 \and more}
\end{abstract}

\section{Introduction}
\label{Introduction1}

Reactions 
\begin{align}
a + A \to s+ F^{*} \to s + b +B
\label{surrogate1}
\end{align}
proceeding through excitation of a resonance in the subsystem $F$ and leading to three charged particles  in the final state play an important role 
in nuclear and atomic physics.  In nuclear physics such reactions are used as surrogate reactions in the Trojan Horse method (THM) \cite{Baur1986,Spitaleri1990,AMM2008,reviewpaper}, which represents a powerful indirect method allowing one to obtain a vital astrophysical information about a binary resonant subreaction
\begin{align}
x + A \to F^{*} \to b+B.
\label{binaryreaction1}
\end{align}
In the THM the initial channel is $a + A$, where $a = (sx)$ is the Trojan Horse (TH) particle, rather than just $x + A$. 
Collision $a+A$ is followed by the transfer reaction $a+ A \to s+ F^{*}$ populating a resonance state $F^{*}$ in the subsystem 
$F=x+A$, which decays to another channel $b + B$. In the THM reaction (\ref{surrogate1}), this results in three particles $b + B + s$ 
in the final state, rather than the two-particle final state $b +B$ in the binary reaction. 

The conditions under which the majority of astrophysical reactions proceed in stellar environments make it difficult or
impossible to measure them under the same conditions in the laboratory. For example, the astrophysical reactions between
charged nuclei occur at energies much lower than the Coulomb barrier, which often makes the cross section of the reaction
too small to measure. This is due to the very small barrier penetration factor from the Coulomb force, which produces
an exponential fall off of the cross section as a function of energy. Typically, reactions that are of interest for nuclear
astrophysics are measured in the laboratory at energies much higher than those relevant to stellar processes.

The indirect THM allows one to bypass the Coulomb barrier issue in the system $x+A$ by using reaction (\ref{surrogate1}) 
rather than the binary reaction (\ref{binaryreaction1}). However, the employing of the surrogate reaction leads to complications of 
both experimental and theoretical nature. Experimentally, coincidence experiments are needed. From the theoretical point of view,
the presence of the third particle $s$ can affect the binary subreaction  (\ref{binaryreaction1}), especially if the particle $s$ 
is charged. In this case, it interacts with the intermediate resonance $F^{*}$ and with the final products $b$ and $B$ via the Coulomb forces, which are very important at low energies and larger charges of participating nuclei. 
For example, the TH reactions induced by collision of the  ${}^{14}{\rm N}$ and ${}^{12}{\rm C}$ nuclei lead to three charged particles in the final state. 
The THM reaction 
\begin{align}
{}^{14}{\rm N} + {}^{12}{\rm C}  \to  d+ {}^{24}{\rm Mg}^{*} \to d + b +B,
\label{14N12Creact1}
\end{align}
where $b= p$ and $\alpha$, $B={}^{23}{\rm Na}$ and ${}^{20}{\rm Ne}$, respectively, can be used to obtain information about the ${}^{12}{\rm C} + {}^{12}{\rm C}$ fusion. This reaction is very important in nuclear astrophysics  \cite{RolfsRodney}. In this reaction the Coulomb effects should have dramatic effect on the cross section.

%The TH reactions induced by collision of the  ${}^{14}{\rm N}$ and ${}^{12}{\rm C}$ nuclei lead to three charged particles in the final state. 
%Recently, the THM reaction 
%\begin{align}
%{}^{14}{\rm N} + {}^{12}{\rm C}  \to  d+ {}^{24}{\rm Mg}^{*} \to d + b +B,
%\label{14N12Creact1}
%\end{align}
%where $b= p$ and $\alpha$, $B={}^{23}{\rm Na}$ and ${}^{20}{\rm Ne}$, respectively, was used to obtain an information about the ${}^{12}{\rm C} + {}^{12}{\rm C}$ fusion \cite{Nature}, which is a very important reaction in nuclear astrophysics  \cite{RolfsRodney}.
%In this reaction the Coulomb effects should have dramatic effect on the cross section.

In atomic physics the surrogate reactions have been used as a tool to investigate the autoionizing states in ion-atom and electron-atom collisions 
\begin{align}
I + A \to I + A^{*} \to I + e+ A^{+},  \nonumber\\
e+ A \to e +A^{*} \to e+ e' + A^{+} .
\label{atsurrogate1}
\end{align}
Even photoionization of the inner atomic shells can be considered as an example of the surrogate reaction:
\begin{align}
\gamma+ A \to e + A^{+*} \to e+e' + A^{++}.
\label{photsurrogate}
\end{align}

A common feature of all these surrogate reactions in atomic and nuclear physics is the presence of three  
charged  particles in the final state. Their post-collision Coulomb interaction may have a crucial impact on the differential cross
sections  \cite{Kuchiev,Senashenko,Godunov,Muk1991}. 

In this paper we use a few-body approach to derive the amplitude and the triple differential cross section of the surrogate reaction
(\ref{surrogate1}) taking into account the Coulomb interaction of the particle $s$ with the resonance $F^{*}$ in the intermediate state 
and the Coulomb interactions of $s$ with $b$ and $B$ in the final state. Our final results reveal a universal effect of the Coulomb interaction  important for both nuclear and atomic surrogate reactions. When applied to atomic reactions, our results coincide with those obtained in \cite{Kuchiev,Senashenko,Godunov}. 

For nuclear surrogate reactions the short-range nuclear interactions should be taken into account alongside the long-range Coulomb interactions. However, the consideration of the nuclear surrogate reaction is simplified due to the fact that the nuclear rescattering effects in the intermediate  $s- F^{*}$ state and within the $s-b$ and $s-B$ pairs in the final state give rise to diagrams that can be treated as a background and disregarded for narrow resonances. 

There is another difference between atomic and nuclear processes. In atomic processes the main interest is related with the effect of the Coulomb interaction on the resonance line shape, the resonance width and shift of the resonance \cite{Kuchiev}. Although in nuclear surrogate reactions these are also of interest, the main problem in the THM is to investigate an energy behavior of  the triply differential cross section which is needed to determine the astrophysical factors \cite{reviewpaper}.
The triply differential cross section derived in this paper allows one to investigate the Coulomb effects on the resonance line shape
both in atomic and nuclear collisions. Moreover, in contrast to \cite{Kuchiev}, we use a general formalism without engaging the eikonal approach. Although in what follows we focus on the derivation of the THM reaction amplitude, the final results are valid also
for atomic surrogate reactions.
 To calculate the energy dependence of the THM triply differential cross section one needs to calculate the differential cross section of the transfer reaction. This transfer reaction is the first part of the resonant surrogate reaction and will be considered in another publication.

To treat the final-state three-body Coulomb effects we use the paper \cite{muk1985}, which was initiated by academician L. D. Faddeev. In this work the formalism of the three-body Coulomb asymptotic states (CAS) was used to calculate the reaction amplitudes with three charged particles in the final state. Both authors are greatly indebted to L.D. Faddeev because few-body physics plays a significant role in our research.

\section{The amplitude of the breakup reaction proceeding through a resonance in the intermediate subsystem}

\subsection{The breakup reaction amplitude in a few-body approach 
%proceeding through resonance in subreaction
}
\label{Breakupampl1}

Let us consider the  surrogate reaction (\ref{surrogate1}), which is the two-step THM reaction.
The difficulty with the analysis of such a reaction stems from the fact that the resonance decays into  the channel 
$b+ B$, which is different from the entry channel $x+A$ of the resonant subreaction. 
As we mentioned earlier, the goal of the THM  is to extract from the TH  reaction \eqref{surrogate1} the  astrophysical factor
for the resonant  rearrangement reaction
%\begin{align}
%x+ A \to F^{*} \to b+B.
%\label
\eqref{binaryreaction1}.
%\end{align}
%
To help the reader follow through all the derivations without complexities caused by the kinematical factors  we consider the collision of spinless particles with the relative orbital angular momenta $l$ in the pairs $x+A$ and $b+B$  set $\,l=0$.  To simplify notation we omit the angular momenta keeping in mind that for all the quantities they are zero.  
The starting expression for the  breakup reaction amplitude in the center-off-mass (c.m.)  of the few-body system can be written as 
\begin{align}
{ M}=  
\blb {\psi_{ {\rm {\bf k}}_{B},{\rm {\bf k}}_{b}  }^{(0)}}
\bl\blb X_{f}\bl{ U}_{0A}\bl\varphi_{a}\,\varphi_{A}\blk \bl
{\psi_{{\rm {\bf k}}_{aA}}^{(0)}}\big>,
\label{Reacampl1}
\end{align}
where $\varphi_{i}$ stands for the bound-state wave function of nucleus $i$, $\,X_{f}= \varphi_{B}\,\varphi_{b}$. The particle $s$  in the THM is a spectator and we can disregard its internal structure and treat it as a  structureless point-like particle.
Wave function  $\,\psi_{ {\rm {\bf k}}_{aA}}^{(0)}$ represents the plane-wave describing the relative motion of the noninteracting  particles $a$ and $A$ in the initial state of the THM reaction,  $\,\psi_{{\rm{\bf k}}_{B},\, {\rm {\bf k}}_{b}}^{(0)}\,$ is the  three-body plane wave of the particles $s,$ $\,b$ and $B$ in the final state, ${\rm {\bf k}}_{i}$ is the momentum of particle $i$. 
%The nucleus $s$ in the final state is a spectator in the THM reaction. 
Note that for the moment we use for charged particles the screened Coulomb potentials. 
The transition operator $\, U_{0A}\,$ corresponds to the breakup reaction from the initial channel $\,a+A\,$ to the final three-body channel $\,s+ b+B$.  The two-fragment partition $\,\alpha + (\beta\,\gamma)\,$ with free particle $\,\alpha\,$ and the bound state $\,(\beta\,\gamma)\,$ is denoted by the free particle index $\,\alpha\,$. 
  
The transition operator $\,U_{0A}\,$ satisfies the equation
\begin{align}
{ U}_{0A}=  V_{f}  +   V_{f}\,G\,{\overline V}_{sx}.
\label{U0A1}
\end{align} 
Here and in what follows we use the following notations:  $V_{\alpha\,\beta} = V_{\alpha\,\beta}^{N} +V_{\alpha\,\beta}^{C}$  is the interaction potential between particles $\alpha$ and 
$\beta$ given by the sum of the nuclear and Coulomb potentials,  ${\overline V}_{\alpha\,\beta}  = V_{\alpha\,\gamma} + V_{\beta\,\gamma}$, and  $\;V_{f}= V_{bB} + V_{sB} + V_{sb}$.

As we mentioned above, the difficulty of the problem is due to the fact that in the TH process the final three-body system $\,s+b+B$, which is formed after the resonance  decay $ F^{*} \to b+B$,  is different from the initial three-body system $s+x+A$  before the resonance $F^{*}$  was formed. This  change  is caused by the rearrangement resonant reaction  
%(\ref{binarysubreaction1}). 
(\ref{binaryreaction1}). 
The full Green's function resolvent  in Eq. \eqref{U0A1} is
\begin{align}
G= \frac{1}{z - {\hat K}_{sF} -V_{sF} - {\hat H}_{F}},
\label{Greenfunction1}
\end{align}
where $\,{\hat K}_{sF}$ is the kinetic energy operator of the $s+ F$ relative motion,  $\,V_{sF}$  is the $s-F$ interaction potential,  $\,{\hat H}_{F}$ is the internal Hamiltonian of the system $\,F= x+A=b+B$.

\begin{figure}[htbp]
\includegraphics[width=0.7\textwidth]{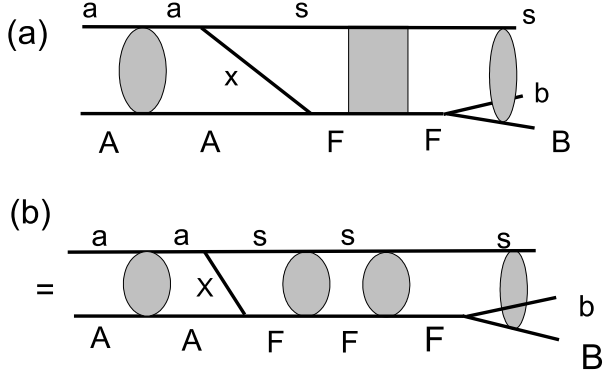}
\caption{The diagrams describing the TH mechanism including the Coulomb interactions in the initial, final and intermediate states. The grey bulb on the left side is the Coulomb $a+A$ scattering in the initial channel described by the Coulomb scattering wave function. The grey rectangle in diagram (a) is the Green's function in the final state describing the propagation of the system $s+F$, where $F$ is the resonance. The grey bulb on the right side describes the intermediate state three-body Coulomb interaction given by the three-body Coulomb wave function. In diagram (b), the Green's function is replaced by its spectral decomposition, which includes the Coulomb scattering wave functions (grey bulbs) describing the Coulomb rescattering of the spectator $s$ in the intermediate state.  }
\label{fig_resdiagram1}
\end{figure}

Our final  goal is to single out the resonance term  in the subsystem $F=x+A=b+B$,  which generates a peak in the triple differential  cross section of the breakup reaction (\ref{surrogate1}), to obtain the reaction amplitude corresponding to the diagrams shown in Fig. \ref{fig_resdiagram1}. 

First we single out  the resonance term  with all the Coulomb rescatterings in the initial, intermediate and final states.
%, which generates the resonance in the triple differential cross section. 
To this end one can rewrite  ${ U}_{0A}\,$ as
\begin{align}
{ U}_{0A}=  V_{f}  +   V_{f}\,G\,{\overline V}_{sx}= V_{f} + (V_{bB}+  {\overline V}_{bB}^{C})\,G\,{\overline V}_{sx}  + {\overline  V}_{bB}^{N}\,G\,{\overline V}_{sx},
\label{U0A2}
\end{align}
where ${\overline V}_{bB}^{NC}= V_{sb}^{NC}   +  V_{sB}^{NC}$.
The resonance term in the subsystem $F$, which can be singled out from  the full Green's function $G,\,$  will be smeared out by the follow up nuclear interactions 
$\,{\overline V}_{bB}^{N}= V_{sb}^{N} + V_{sB}^{N}.\,$   Hence the term  $ {\overline  V}_{bB}^{N}\,G\,{\overline V}_{sx}$  does not produce the resonance peak in the TH reaction amplitude generated by the resonance in the subsystem $F\,$  and can be treated as a background. That is why the  only term  in Eq. (\ref{U0A2}), which is responsible for  a resonance behavior of the TH reaction amplitude caused  by the resonance in the subsystem $F,\,$  is
\begin{align}
 { U}_{0A}^{{\rm R}} =  (V_{bB}+ {\overline V}_{bB}^{C} )\,G\,{\overline V}_{sx} .
 \label{U0AR3}
 \end{align}
 Note that the Coulomb potentials $\, V_{sB}^{C}$ and $\,V_{sb}^{C}$ do not smear out the resonance in the subsystem $\,F,\,$ which can be singled out from $\,G$ \cite{blokh84}.
  
 Substituting  ${ U}_{0A}^{{\rm R}}$  into the matrix element (\ref{Reacampl1})  instead of $\,{ U}_{0A}\,$  leads to
\begin{align}
{ M}' &= 
\big <{\psi_{ {\rm {\bf k}}_{B},{\rm {\bf k}}_{b}}^{(0)}}\big|
\big<{X_{f}}\big|{(V_{bB}+  {\overline V}_{bB}^{C}\big) \,G\,{\overline V}_{sx}}\big|\varphi_{a}\,\varphi_{A}\blk\bl
{  \psi_{{\rm {\bf k}}_{aA}}^{(0)}}\big>
\label{Min1}   \\
&= 
\blb{{ \Phi}_{ {\rm {\bf k}}_{B},{\rm {\bf k}}_{b}}^{C(-)}}
\bl\blb{X_f}\bl{V_{bB}\,G\,{\overline V}_{sx}}\bl\varphi_{a}\,\varphi_{A}\blk\bl
{ \psi_{{\rm {\bf k}}_{aA}}^{(0)}}\blk,
\label{Min2}    
\end{align}
where ${\overline V}_{bB}^{C}= V_{sB}^{C}  + V_{sb}^{C}$.

To obtain Eq. (\ref{Min2})  from Eq. (\ref{Min1})  the two-potential formula is applied.   Application of this formula allows one to subtract the sum of the Coulomb potentials
$ V_{sB}^{C}  + V_{sb}^{C}$ from the transition operator $V_{bB}+  V_{sB}^{C}  + V_{sb}^{C}$ in Eq. (\ref{Min1})  with simultaneous replacement of the final-state  
three-body plane wave  $\psi_{ {\rm {\bf k}}_{B},{\rm {\bf k}}_{b}  }^{(0)}$  by the three-body Coulomb wave function $\,{  \Phi}_{ {\rm {\bf k}}_{B},{\rm {\bf k}}_{b}}^{C(-)}$  in the bra state. The latter is a solution of the Schr\"odinger equation
\begin{align}
{  \Phi}_{ {\rm {\bf k}}_{B},{\rm {\bf k}}_{b}}^{C(-)*}\big( E_{f} - {\overleftarrow {\hat K}}_{sbB} -  V_{sB}^{C} - V_{sb}^{C} \big) = 0,
\label{3BCSheq1}
\end{align}
with the incoming-wave boundary condition. In this equation the Coulomb interaction in the pair $\,b+B\,$ is absent. ${\overleftarrow {\hat K}}_{sbB}\,$, which acts to the left, is the total kinetic energy operator  of  the three-body system $s+b+B$ and   $\,E_{f}$ is the total kinetic energy of the system $\,s+b+B$  in the  c.m.  system. 

Now rewriting the potential $\,{\overline V}_{sx}\,$ as $\,{\overline V}_{sx} = {\overline V}_{sx}^{N} + {\overline V}_{sx}^{C} $   and applying the two-potential formula, we reduce Eq. (\ref{Min2}) to
 \begin{align}                                     
{ M}'=  
%<{\overline \Phi}_{bB}^{C(-)}\,\,X_{f}\,\big| V_{bB}\,G\,{\overline V}_{sx}^{N}\,\big| \varphi_{a}\,\varphi_{A}\,\Psi_{{\rm {\bf k}}_{aA}}^{C(+)} >.\\
\blb{{  \Phi}_{ {\rm {\bf k}}_{B},{\rm {\bf k}}_{b}}^{C(-)}}\bl\blb{X_f}\bl{V_{bB}\,G\,{\overline V}_{sx}^{N}}\big|\varphi_{a}\,\varphi_{A}\blk\bl
{\Psi_{{\rm {\bf k}}_{aA}}^{C(+)} }\blk.
\label{Mf1}
\end{align} 
Here, $\,\Psi_{{\rm {\bf k}}_{aA}}^{C(+)}\,$ is the $\,a+A\,$  Coulomb scattering wave function, which satisfies  the two-body Schr\"odinger equation with  the Coulomb potential  ${\overline V}_{sx}^{C}= V_{sA}^{C} + V_{xA}^{C}$ with the outgoing-wave boundary condition.

It is important to explain again why we are taking into account only the Coulomb distortions in the initial and final states rather than the Coulomb plus nuclear distortions. 
First, it will be shown that the Coulomb rescatterings  transform the resonance pole into the branching point without smearing out the resonance peak.
Second, in the THM, as used currently \cite{reviewpaper}, only the energy dependence of the  extracted astrophysical factor  is measured  while its absolute value 
is determined by the normalization to the available direct data.  The Coulomb rescatterings in the initial, intermediate and final states, in contrast to the nuclear distortions, can significantly modify  the energy dependence of the THM differential cross section for the reactions at sub-Coulomb and near the Coulomb barrier energies and, therefore, must  be taken into account. 

The TH reaction proceeds as a two-step mechanism. On the first step the particle $x$ is transferred from the bound state $a=(s\,x)$ to the  nucleus $A$ forming a resonance $\,F^{*}= x+A$. It is assumed in the THM that $x$ is transferred in the ground state. To provide the transfer of particle $x$ in the ground state we introduce the projector $\,P_{x}=  \sum \bl\varphi_{x} \blk \blb \varphi_{x}\bl$, where $\,\sum\,$ denotes the sum over the bound states and the integration over the  continuum of the nucleus $\,x$. To keep $\,x\,$ in the ground state we leave in the projector only the term $\bl\varphi_{x} \blk \blb \varphi_{x} \bl$, where $\varphi_{x}$ is the bound-state wave function in the ground state. Then we can get from Eq. (\ref{Mf1})
\begin{align}                                     
{ M}'=  
%<{\overline \Phi}_{bB}^{C(-)}\,\,X_{f}\,\big| V_{bB}\,G\,{\overline V}_{sx}^{N}\,\big| \varphi_{a}\,\varphi_{A}\,\Psi_{{\rm {\bf k}}_{aA}}^{C(+)} >.\\
\blb{{  \Phi}_{ {\rm {\bf k}}_{B},{\rm {\bf k}}_{b}}^{C(-)}}\bl\blb{X_f}\bl{V_{bB}\,G\,{\overline V}_{sx}^{N}}\bl X_{i}\blk \bl I_{x}^{a}\,
{\Psi_{{\rm {\bf k}}_{aA}}^{C(+)} }\blk,
\label{Mf11}
\end{align} 
where $X_{i}= \varphi_{x}\,\varphi_{A}$. We also introduce the overlap function of the bound-state wave functions of nuclei $a$ and $x$:
\begin{align}
I_{x}^{a}(r_{sx}) = \blb \varphi_{x}(\xi_{x}) \bl \varphi_{a}(\xi_x; r_{sx}) \blk 
%\bl_{xi_{x}}
,
\label{ovfIax1}
\end{align}
which is the projection of the bound-state wave function $\,\varphi_{a}\,$ on the two-body channel $\,s+x$.
The integration in the matrix element is carried over the internal coordinates $\,\xi_{x}\,$ of nucleus $x$. $\,r_{sx}\,$ is the radius connecting the c.m. of nuclei $\,s\,$ and $\,x$. Since we assume that the relative  $\,s-x\,$ orbital angular  momentum  in the bound  state $\,a=(sx)\,$ equals zero the overlap function depends on $\,r_{sx}\,$ rather than on $\,{\rm {\bf r}}_{sx}.\,$  Note that we also neglected the internal degrees of the spectator $\,s$. Otherwise, we need to include in the bra state of the matrix element in Eq. (\ref{ovfIax1}) the bound-state wave function of $s$ and integrate over $\,\xi_{s}$.

Now it is time to single out from the Green's function $G$ the resonance in the subsystem $\,F= x+A\,$.  To this end  one can rewrite 
\begin{align}
 G= {G}_{s}\,\big(1 +  {\overline V}_{xA}^{N}\,G \big),
\label{GGs1}
\end{align}
where ${\overline V}_{xA}^{N}= V_{sx}^{N} + V_{sA}^{N}$ and 
\begin{align}
{G}_{s}(z) = \frac{1}{z - {\hat K}_{sF}  -  {\hat H}_{F} -  {\overline V}_{sF}^{C}}.
\label{tildeGs1}
\end{align}
% $\;{\hat H}_{sxA}= {\hat H}_{s} + {\hat H}_{x}+ {\hat H}_{A}$ is the sum of the internal Hamiltonians of the final-state nuclei $s,\;x,\,A$.
Note that ${\overline V}_{sF}^{C}={\overline V}_{xA}^{C}=V_{sx}^{C}+V_{sA}^{C}$.

Substituting Eq. (\ref{GGs1}) into Eq. (\ref{Mf1})  one gets
 \begin{align}  
 { M}=  \blb{{  \Phi}_{ {\rm {\bf k}}_{B},{\rm {\bf k}}_{b}}^{C(-)}}\bl\blb{X_f}\bl{V_{bB}\, G_{s}\,{ {\tilde U}}_{sA}}\bl X_{i}\blk \bl I_{x}^{a}\,
{\Psi_{{\rm {\bf k}}_{aA}}^{C(+)} }\blk.
\label{Mf2}
\end{align} 
Here the transition operator ${ {\tilde U}}_{sA}$  is
\begin{align}
{ {\tilde U}}_{sA}  = {\overline V}_{sx}^{N} + {\overline V}_{xA}^{N}\,G\,{\overline V}_{sx}^{N}.
\label{tildeUSA1}
\end{align}
It is important to underscore that when deriving Eq. (\ref{Mf2}) we made only one approximation by inserting $\bl\varphi_{x} \blk \blb \varphi_{x}\bl$. Now we introduce the second approximation by replacing ${\overline V}_{sF}^{C}$  in Eq. \eqref{tildeGs1} with $U_{sF}^{C}$: 
\begin{align} 
{G}_{s} (z)= \frac{1}{z - {\hat K}_{sF}  -  {\hat H}_{F} -  {\overline V}_{sF}^{C} }
\approx    \frac{1}{z - {\hat K}_{sF}  -  {\hat H}_{F} - U_{sF}^{C} }   .
\label{GstildeGs1}
\end{align}
This will allow us to simplify the  spectral decomposition of $G_{s}$.
Here $U_{sF}^{C}$ is the channel  Coulomb potential describing the  interaction between the c.m. of nuclei $s$ and $F$. 

Then the reaction amplitude ${ M}$ takes the form 
 \begin{align}                                     
{ M}'=  
\blb {{  \Phi}_{ {\rm {\bf k}}_{B},{\rm {\bf k}}_{b}}^{C(-)}} \bl
\blb {X_f} \bl V_{bB}\,G_{s}\, {\tilde U}_{sA}\bl {X_i} \blk \bl I_{x}^{a}
{\Psi_{{\rm {\bf k}}_{aA}}^{C(+)}} \blk.                              
\label{Mf3}
\end{align} 

To single out the first step of the THM reaction, the particle $x$ transfer reaction $\,a+ A \to s +F^{*}\,$  we introduce the projector $\,P_{i}= \sum \bl X_{i} \blk \blb X_{i} \bl,\,$  where  $\,\sum\,$  stands for the sum over the bound states and the integration over the  continuum of the nuclei $\,x\,$ and $\,A$. We keep in this projector only one term, $\,\bl X_{i} \blk \blb X_{i} \bl.\,$ This  will ensure that the intermediate state after the transition operator $\,{ {\tilde U}}_{sA}\,$ is $\,s+ x+ A\,$, where all three particles are  in their respective ground states. 
 
Also we  introduce the second projector operator $\;{ P}_{f}= \sum \bl X_{f} \blk \blb X_{f} \bl.\,$  Here,  $\,\sum\,$  denotes the sum over the bound states and the integration over the  continuum of the nuclei $b$ and $B$.  Again, we leave in this projector only the term,  $
%\; { P}_{f} \approx 
\bl X_{f} \blk \blb X_{f} \bl$,  where $X_{f} =  \varphi_{b}\,\varphi_{B}$ assuming that resonance decay products are formed in the ground states. These ground states can be replaced by excited states.

Inserting the projectors ${X_{f}}$ and  ${X_{i}}$ into Eq. (\ref{Mf3})  before  and after $G_{s}$, respectively, one gets:
 \begin{align}                                     
{ M}' &= 
\blb {{  \Phi}_{ {\rm {\bf k}}_{B},{\rm {\bf k}}_{b}}^{C(-)}} \bl \blb {X_{f}} \bl {V_{bB}} \bl {X_{f}} \blk
  \blb   {X_{f}} \bl { G_{s}} \bl {X_{i}}   \blk
   \blb {X_{i}} \bl {{ {\tilde U}}_{sA}} \bl {X_i} \blk  \bl
  I_{x}^{a}\,{\Psi_{{\rm {\bf k}}_{aA}}^{C(+)}} \blk             \nonumber\\
&= 
\blb{{  \Phi}_{ {\rm {\bf k}}_{B},{\rm {\bf k}}_{b}}^{C(-)}} \bl
{{\tilde V}_{bB} \bl
  \blb  {X_{f}} \bl  { G_{s}} \bl {X_{i}} \blk \bl  {\cal U}_{sA}} \bl I_{x}^{a}\,
{\Psi_{{\rm {\bf k}}_{aA}}^{C(+)}} \blk,         
\label{Mfpgrst2}
\end{align} 
where
%\begin{align}
%{\cal M}^{\rm BR}= 
%<  X_{i} \big| \,{\cal {\tilde U}}_{sA}\,\big| \varphi_{a}\,\varphi_{A}\,\Psi_{{\rm {\bf k}}_{aA}}^{C(+)} >         
%\label{MBRij1}
%\end{align}
%is the breakup reaction amplitude  of the reaction $a+A \to s + x + A$. We also 
we introduced  short-hand notations $\,{\tilde V}_{bB}\,= 
\blb {X_f}\big|{V_{bB}}\bl {X_f}\blk $ and ${\cal U}_{sA}= \blb{X_{i}}\big|{{ {\tilde U}}_{sA}}\big|{X_i}\blk$. We assume that the potential $V_{bB}$  is spin-independent.
 
 \subsection{Spectral decomposition of  the two-channel Green's function}
 \label{spectraldecGrfunct1}

To single out a resonance state in the intermediate subsystem $\,F\,$ one can introduce the spectral decomposition  of  $\,G_{s} (z)$:
\begin{align}
\blb X_{f} \big| G_{s}  \bl  \,X_{i}>\, =&\,  \sum_{n}\,\int\,\frac{{\rm d} {\rm {\bf k}}_{sF}}{(2\,\pi)^{3}}\,\frac{  \bl  I_{f}^{F_{n}}\,\Psi_{{\rm {\bf k}}_{sF} } ^{(-)} \blk \blb \Psi_{{\rm {\bf k}}_{sF} }^{(-)} I_{i}^{F_{n}} \bl }
{E_{aA} + Q_{n} - k_{sF}^{2}/(2\,\mu_{sF})   + i0 }                                                     \nonumber\\
&+ \,\int\, \frac{{\rm d} {\rm {\bf k}}_{bB}}{(2\,\pi)^{3}}\,\frac{{\rm d} {\rm {\bf k}}_{sF}}{(2\,\pi)^{3}}\,  \frac{ \bl \Psi_{{\rm {\bf k}}_{bB};f}^{(-)}
\, \Psi_{{\rm {\bf k}}_{sF} }^{C(-)} \blk  \blb \Psi_{ {\rm {\bf k}}_{sF} }^{C(-)}\,\Psi_{{\rm{\bf k}}_{bB};i}^{(-)}\bl }
{  E_{aA} - \varepsilon_{a} +Q_{if} -  k_{bB}^{2}/(2\,\mu_{bB}) - k_{sF}^{2}/(2\,\mu_{sF})  + i0 }.
\label{spectrtildeGs1}
\end{align} 
Here we again use the notion of the overlap function $I_{f}^{F_{n}}(r_{bB}) = \blb \varphi_{B}(\xi_{B})\,\varphi_{b}(\xi_{b}) \bl \varphi_{n}(\xi_{B},\xi_{b}; r_{bB}) \blk $, which is the projection of the $n$-th bound-state of the many-body wave function $\varphi_{n}$  of $\,F$  on $X_{f}$. Integration in the 
%subscript $f$ in the 
matrix element 
%means that the integration 
is carried out over all the internal coordinates $\xi_{b}$ and $\xi_{B}$ of nuclei $b$ and $B$.
%, respectively. 
Hence, $I_{f}^{F_{n}}$ depends only on $r_{bB}$ (for non-zero orbital angular momenta it depends on ${\rm {\bf r}}_{bB}$).
Similar meaning has the second overlap function $I_{i}^{F_{n}}(r_{xA})= \blb \varphi_{n}(\xi_{A},\xi_{x}; r_{xA}) \bl \varphi_{x}(\xi_{x})\,\varphi_{A}(\xi_{A}) \blk $ introduced in Eq. \eqref{spectrtildeGs1}.

Usually the overlap functions are determined for bound states. But here we also introduce the overlap functions for 
the continuum states:
\begin{align}
\Psi_{ {\rm {\bf k}}_{bB};f}^{(-)}({\rm {\bf r}}_{bB})  &=  \blb \varphi_{B}(\xi_{B})\,\varphi_{b}(\xi_{b}) \bl  \Psi_{{\rm {\bf k}}_{bB}}^{(-)}(\xi_{B},\xi_{b}; {\rm {\bf r}}_{bB}) \blk        \\
%\Psi_{- {\rm {\bf k}}_{bB};i}^{(+)}({\rm {\bf r}}_{xA}) & =  \blb \varphi_{A}(\xi_{A})\,\varphi_{x}(\xi_{x}) \bl  \Psi_{-{\rm {\bf k}}_{bB}}^{(+)}(\xi_{A},\xi_{x}; {\rm {\bf r}}_{xA}) \blk \\
\Psi_{ {\rm {\bf k}}_{bB};i}^{(-)*}({\rm {\bf r}}_{xA}) & =  \blb  \Psi_{{\rm {\bf k}}_{bB}}^{(-)}(\xi_{A},\xi_{x}; {\rm {\bf r}}_{xA}) \bl \varphi_{A}(\xi_{A})\,\varphi_{x}(\xi_{x}) \blk 
\label{ovfPsif1}
\end{align}
 are the projections of the  wave function $\Psi_{{\rm {\bf k}}_{bB}}^{(-)}$ 
 %and  $\Psi_{-{\rm {\bf k}}_{bB}}^{(+)}$ 
 of the system $F$ in the continuum on $X_{f}$ and $X_{i}$, respectively. 
 %The subscripts $f$  and $i$ in the matrix elements has the same meaning as for the matrix elements with the bound states. 
 %We took into account that $\Psi_{{\rm {\bf k}}_{bB}}^{(-)*}= \Psi_{-{\rm {\bf k}}_{bB}}^{(+)}$. 
 We  assume that the continuum wave function $\Psi_{{\rm {\bf k}}_{bB}}^{(-)}$ has the incident wave in the channel $f=b+B$  with ${\rm {\bf k}}_{bB}$ being the $b+B$ relative momentum.

Also in Eq. \eqref{spectrtildeGs1},  $\;E_{aA}$ is the $a-A$ relative kinetic energy,  $Q_{n}= m_{a} + m_{A} - m_{s} - m_{F_{n}}= \varepsilon_{F_{n}} - \varepsilon_{a}$,  $\,\varepsilon_{F_{n}}\, = m_{x} + m_{A} - m_{F_{n}}$  is   the binding energy of the bound state $\,F_{n}\,$ for the virtual decay $\,F_{n} \to x+A$,  $\,\Psi_{{\rm {\bf k}}_{sF} }^{C(-)}\,$ is the Coulomb scattering wave function of particles $s$ and $F$ with the relative momentum  $\,{\rm {\bf k}}_{sF}$, $\,\mu_{sF}\,$  is the reduced mass of particles $s$ and $F$, $\;\varepsilon_{a} = m_{s} + m_{x} - m_{a}$ is the binding energy  of $a$, $\,m_{i}$  is the mass of particle $i$,  $\,E_{aA}-\varepsilon_{a} + Q_{if}$  is the total kinetic energy of the three-body system   $s+b +B$,   $\,Q_{if}= m_{x} + m_{A} - m_{b} - m_{B}$,  $m_{i}$ is the mass of particles $i$,  $i=x+A$ and $f=b+B$  are the initial and final channels of the binary subreaction $x+ A \to b+B$.

 In the external region  (the reason why it is enough to consider only the external region is explained in \cite{muk2011})  the wave function  $\, \Psi_{{\rm {\bf k}}_{bB}}^{(-)}$ with the incident wave  in the channel $f=b+B$   becomes an external multichannel scattering wave function  \cite{muk2011,lanethomas}:
 \begin{align}
&\Psi _{{\rm {\bf k}}_{bB}}^{(-)}({\rm {\bf r}}_{bB}) = \,\,-i\frac{1}{2\,k_{bB}}\sum\limits_c {\sqrt {\frac{{{v_f}}}{{{v_c}}}} } \frac{1}{{{r_c}}}{{ X}_c}\, \big [I^*(k_{bB} ,\,r_{bB})\,\delta _{cf} - \,\,S^*_{cf}\,O^*({k_c},{r_c}) \big] .
\label{multuchPsi1} 
\end{align}
Note that on the left-hand-side $\Psi _{{\rm {\bf k}}_{bB}}^{(-)}({\rm {\bf r}}_{bB})$ depends on the vectors
${\rm {\bf k}}_{bB}$ and ${\rm {\bf r}}_{bB}$ while the right-hand-side depends only on the scalars.  It is because on the right-hands-side we take into account only the $l=0$ partial wave.  
 
 We recall that $f$ stands for the channel $b+B$.  The sum over $c$ is taken over all open final channels $\,c\,$ coupled with the initial channel $f$.  $\,X_{c}$ stands for the product  of the bound-state wave functions of the fragments in the channel $c$,  $\,v_{c}\,$ is the relative velocity of the nuclei in the channel $c$, $S_{c\,f}$ is the scattering $S$ matrix for the transition $\, f \to  c$, $\;O(k_{c},r_{c})$ 
 %and   $O(k_{c},r_{c})\,$  are 
 is the Coulomb Jost singular solution of the Schr\"odinger equation  with the 
 %ingoing and 
 outgoing-wave asymptotic behavior.
 %correspondingly. 

 In the case under consideration we consider only two coupled channels, $\,i= x+A$ and  $\,f=b+B$.  
    In the external region the channels are decoupled  and the overlap function  $\Psi_{{\rm {\bf k}}_{bB}; i}^{(-)}$ 
    is written as
  \begin{align}
  \Psi_{{\rm {\bf k}}_{bB};i}^{(-)*}({\rm {\bf r}}_{xA}) = -i\frac{1}{{{2\,k_{bB}}{\,r_{xA}}}} {\sqrt {\frac{{{\mu_{xA}\,k_{bB}}}}{{{\mu_{bB}\,k_{xA}}}}} }\,S_{  f\, i}\,O({k_{xA}},{r_{xA}}). 
   \label{overlapfunctionsxA1}
 \end{align} 
 Equation (\ref{overlapfunctionsxA1})  determines the projection of the external two-channel wave function $\Psi_{F}^{(-)}$, which has an incident wave in the channel $\,f=b+B$, onto the channel $\,i=x+A$.  
 The second overlap function takes the form
 \begin{align}
&\Psi_{{\rm {\bf k}}_{bB};f}^{(-)}({\rm {\bf r}}_{bB}) =  \, -i\frac{{1 }}{{{2\,k_{bB}}{r_{bB}}}}\,\big[ I^{*}({k_{bB}},{r_{bB}})  - \,\,S_{f\,f}^{*}\,O^{*}({k_{bB}},{r_{bB}}) \big],
\label{Psiff1}
\end{align}
where $I^{*}({k_{bB}},{r_{bB}})=  O({k_{bB}},{r_{bB}})$.    
 
We denote 
%the latter as $ \blb  X_{f} \big|{ G}_{s}^{\rm cont} \bl X_{i} \blk$:
%Then 
the second (continuum) term in Eq. (\ref{spectrtildeGs1}) as $ \blb  X_{f} \big|{ G}_{s}^{\rm cont} \bl X_{i} \blk$:
% takes the form 
  \begin{align}
  \blb  X_{f} \big|{ G}_{s}^{\rm cont} \bl X_{i} \blk = &  \int   \frac{{\rm d} {\rm {\bf k}}_{bB}}{(2 \pi)^{3}} \frac{{\rm d} {\rm {\bf k}}_{sF}}{(2 \pi)^{3}}   \frac{ \bl \Psi_{ {\rm {\bf k}}_{bB};f}^{(-)} 
  \Psi_{{\rm {\bf k}}_{sF} }^{C(-)} \blk \blb \Psi_{{\rm {\bf k}}_{sF} }^{C(-)}  \Psi_{{\rm {\bf k}}_{bB};i}^{(-)} \bl  }
{ E_{aA} - \varepsilon_{a} + Q_{if} - k_{bB}^{2}/(2 \mu_{bB})  - k_{sF}^{2}/(2 \mu_{sF})  + i0 }. 
\label{spectrtildeGscont1}
\end{align}
 The resonant term corresponding to the subsystem  $F$  can be singled out from Eq. (\ref{spectrtildeGscont1}). 
To this end
  \begin{enumerate}
  \item  We first perform the integration over the solid angle $\Omega_{{\rm {\bf { k}}}_{bB}}$. 
  \item The $S$-matrix element  $S_{i\,f}$  has a resonance pole on the second Riemann sheet at the $b-B$ relative energy $E_{{\rm R}(bB)} =E_{0(bB)} - i\Gamma/2$,  where $\Gamma$ is the total resonance width.  In the momentum plane this resonance pole  occurs at the $b-B$ relative momentum  $k_{{\rm R}(bB)}= k_{0(bB)} - i\,k_{I(bB)}$. We assume that  the resonance is narrow:  $\Gamma  << E_{0(bB)}$  or $k_{I(0)} << k_{0(bB)}$. 
   \item  When $k_{bB}  \to k_{{\rm R}(bB)}$  the integration contour over $k_{bB}$  moves down to  the fourth quadrant  pinching the contour to the pole at $k_{{\rm R}(bB)}$. Taking the  residue at the pole  $\,E_{bB}= E_{{\rm R}(bB)}\,$ one can single out the contribution to  $ \blb X_{f} \big|{ G}_{s}^{\rm cont} (z) \bl X_{i} \blk\,$ from the resonance term in the subsystem $\,F$. 
  \end{enumerate}
  
  Now we can proceed to the practical realization of the outlined  scheme.  First, substituting Eqs. (\ref{overlapfunctionsxA1})  and (\ref{Psiff1}) into  Eq. (\ref{spectrtildeGscont1}) and integrating over the solid angle $\Omega_{{\rm {\bf k}}_{bB}}$  we get
 \begin{align}
\blb X_{f} \big| G_{s}^{\rm cont} \bl X_{i} \blk =  &
\,-\frac{\pi}{ {r_{bB}\,r_{xA}}}\,\int\limits_{0}^{\infty}\,\frac{{\rm d} {k}_{bB}}{(2\,\pi)^{3}}\,\frac{{\rm d} {\rm {\bf k}}_{sF}}{(2\,\pi)^{3}}\, {\sqrt {\frac{{{\mu_{xA}\,k_{bB}}}}{{\mu_{bB}\,{k_{xA}}}}} }                                                \nonumber\\
& \times  \, \frac{ \, \big[ O({k_{bB}},{r_{bB}})  - \,\,S_{f\,f}^{*}\,O^{*}({k_{bB}},{r_{bB}}) \big]                 
\,  \bl  \Psi_{{\rm {\bf k}}_{sF} }^{C(-)} \blk \blb \Psi_{{\rm {\bf k}}_{sF} }^{C(-)}\Big| \, S_{f\,i}\, {O({k_{xA}},{r_{xA}}) }}
{ E_{aA} - \varepsilon_{a} + Q_{if} - k_{bB}^{2}/(2\,\mu_{bB})  - k_{sF}^{2}/(2\,\mu_{sF})  + i0 }.
\label{spectrtildeGscont21}
\end{align}             
 We also took into account  that at real $k_{bB},$     $\, I^{*}({k_{bB}},{r_{bB}})= O({k_{bB}},{r_{bB}})$.
  
 The reaction $\,S\,$-matrix element  $\,S_{f\,i} $   has a resonance pole in the fourth quadrant on the second  sheet of the energy $E_{bB}$ plane:
\begin{align}
S_{f\,i }  \stackrel{E_{bB} \to E_{{\rm R}(bB)}}{=}    e^{i[\delta^{p}( k_{0(bB)})  + \delta^{p}(k_{0({xA})}) ]}\,\frac{\Gamma_{bB}^{1/2}\,\Gamma_{xA}^{1/2} }{ E_{0(bB)} - E_{bB} -i\,\Gamma/2}.
\label{Sfi11}
\end{align}   
$\,\Gamma= \Gamma_{xA} + \Gamma_{bB}$ is the total resonance width, $\,\Gamma_{xA }\,$ and $\,\Gamma_{bB}\,$ are the partial resonance widths for the decay of the resonance into the channels $i=x+A$ and $f=b+B$, respectively.  $\,\delta^{p}( k_{0(bB)})\,$ and $\,\delta^{p}(k_{0(xA)})\,$ are the potential scattering phase shifts in the channels
 $f$ and $i$. The imaginary parts of the resonance momenta in their arguments are neglected  because we consider a narrow resonance.

 Note that  the complex conjugated  elastic scattering matrix $S$-matrix element   $\,S_{f\,f}^{*}$  does not have a pole in the fourth quadrant on the second  sheet of the energy $E_{bB}$ plane:
  \begin{align}
S_{f\,f}^{*}=  e^{-2\,i\,\delta^{p}( k_{0(bB)})}\,\frac{ E_{0(bB)} -  E_{bB} + i\Gamma/2 - i\Gamma_{bB}}
{E_{0(bB)} - E_{bB} +  i\,\Gamma/2}   \stackrel{E_{bB} \to E_{{\rm R}(bB)}}{=}   e^{-2i{\delta^{p}(k_{0(bB)})} }\, \frac{ \Gamma_{xA}}{\Gamma }.
\label{Sff11}
\end{align}     
 
 At  $\,k_{bB} \to k_{{\rm R}(bB)}$   the integration contour over $k_{bB}$ moves down  to the fourth quadrant  contour pinching  it to  the pole  $k_{bB}= k_{{\rm R}(bB)}\,$  ($k_{xA}= k_{{\rm R}(xA)}$). Note that this pole corresponds to the pole $E_{bB}= E_{{\rm R}(bB)}$ in the fourth quadrant of the second energy sheet.  
 Substituting Eqs.  (\ref{Sfi11})  and (\ref{Sff11}) and taking the residue of $S_{f \,i}$  in the pole  we get the resonance term:
 \begin{align}
\blb X_{f} \big|G_{s}^{\rm cont} \bl  X_{i} \blk =& - e^{i[\delta^{p}( k_{0(bB)})  + \delta^{p}(k_{0(xA)}) ]}
\frac{i}{4\,\pi}\,  \sqrt {\frac{\mu_{ bB}\,\mu_{xA}}{k_{{\rm R}(bB)}\,k_{{\rm R}(xA)} }}\,\int\,\frac{{\rm d} {\rm {\bf k}}_{sF}}{(2\,\pi)^{3}}\,                                                \nonumber\\
& \times    \,\big[ O({k_{{\rm R}(bB)}},{r_{bB}}) -   e^{-2\,i\,\delta^{p}( k_{0({(bB)})})}\, ({\Gamma_{(xA)} }/{ \Gamma })\,O^{*}({k_{{\rm R}(bB)}},{r_{bB}}) \big]      
\nonumber \\ & \times          
\frac{\bl \Psi_{{\rm {\bf k}}_{sF} }^{C(-)} \blk \blb \Psi_{{\rm {\bf k}}_{sF} }^{C(-)} \, \bl \Gamma_{(bB)}^{1/2}\,\Gamma_{(xA) }^{1/2}\,O({k_{{\rm R}(xA)}},{r_{xA}}) }
{ E_{aA} - \varepsilon_{a} + Q_{if} - E_{{\rm R}(bB)}  - k_{sF}^{2}/(2\,\mu_{sF}) }          
%\nonumber \\ & 
+  \blb X_{f} \big|G_{\rm NR} \bl X_{i} \blk.
\label{spectrtildeGsR21}
\end{align}  
 Here $k_{{\rm R}(bB)}$ and $k_{{\rm R}(xA)}$ are the resonance momenta in the channels $f$ and $i$. 
The term $ \blb X_{f} \big|G_{\rm NR}\big| X_{i} \blk$ is the nonresonant term  and in what follows we neglect it.
%in Eq. (\ref{spectrtildeGsR21}).
Usually,  in the THM,  ${ \Gamma_{(xA)}}/{\Gamma} << 1$. Besides,  $O^{*}({k_{{\rm R}(bB)}},{r_{bB}}) $ exponentially decreases when $r_{bB} \to \infty$. Therefore, neglecting the term   $  ({\Gamma_{(xA)}   }/{ \Gamma})\, O^{*}({k_{{\rm R}(bB)}},{r_{bB}})$  we get the desired spectral decomposition of $G_{s}$ for two coupled channels:
  \begin{align}
& \blb  X_{f} \big|  G_{s}^{{\rm R}} \bl X_{i} \blk =- \frac{i}{4\,\pi}\,\int\,\frac{{\rm d} {\rm {\bf k}}_{sF}}{(2\,\pi)^{3}}\,                                                
 \frac{ {\, \bl  \phi_{{\rm R}(bB)}(r_{bB})\, \Psi_{{\rm {\bf k}}_{sF} }^{C(-)} \blk \blb \Psi_{{\rm {\bf k}}_{sF} }^{C(-)} \,  {\tilde \phi}_{{\rm R}(xA)}(r_{xA})} \bl  }
{ E_{aA} - \varepsilon_{a} + Q_{if} - E_{{\rm R}(bB)}  - k_{sF}^{2}/(2\,\mu_{sF}) },
\label{spectrtildeGsRtf1}
\end{align} 
where  
\begin{align}
& {\tilde \phi}_{{\rm R}(xA)}(r_{xA}) = e^{-i\,\delta^{p}(k_{0(xA)})}\,\sqrt{    \frac{\mu_{xA}}{k_{{\rm R}(xA)} }\,\Gamma_{xA} }\,\frac{O^{*}({k_{{\rm R}(xA)}},{r_{xA}})}{r_{xA}},  \nonumber\\
& \phi_{{\rm R}(bB)}(r_{bB}) =e^{i\,\delta^{p}(k_{0(bB)})}\, \sqrt{    \frac{\mu_{bB}}{k_{{\rm R}(bB)} }\,\Gamma_{bB} }\,\frac{O({k_{{\rm R}(bB)}},{r_{bB}})}{r_{bB}} 
\label{Gamowwfi1}
\end{align}
are the Gamow resonant wave functions in channels $i$ and $f$. Note that $\, {\tilde \phi}_{{\rm R}(xA)}(r_{xA})\,$  is the Gamow wave function from the dual basis.

\section{Amplitude of $a+ A \to s+b+B$ reaction 
%with three charged particles in the final state 
proceeding through  a resonance in the intermediate binary subsystem
}
\label{BRresampl1}

After deriving the expression for  the resonance term  in the  spectral decomposition of   $\blb X_{f} \big| G_{s}^{{\rm R}} (z) \big| X_{i}\blk$  we can substitute it into Eq. (\ref{Mfpgrst2})  and  derive an equation for the amplitude of the reaction  $a+ A  \to s+b+B$  with three charged particles in the final state, proceeding through an intermediate resonance in the subsystem  $F=x+ A=b+B$.  

Now we are in position to write down the TH reaction amplitude proceeding through the resonance in the intermediate binary subsystem. As mentioned above, the TH reaction amplitude is described by the two-step process: the first step is the transfer reaction populating the resonance state 
$a+ A \to s +F^{*}$ and the second step is the decay of the resonance into two-fragment channel $F^{*} \to b+B$   leading to the formation of the three-body final state, $s+ b+ B$.  We derive below the expression for the TH reaction amplitude taking into account the Coulomb interactions in the intermediate and final states. 

Substituting Eq. (\ref{spectrtildeGsRtf1})  into Eq. (\ref{Mfpgrst2}) and writing it in the momentum representation  one gets
%\begin{widetext}                                     
\begin{align}                                     
%&{ M}'= - \frac{i\,\mu_{sF}}{2\,\pi} \,\int\,\frac{{\rm d} {\rm {\bf k}}_{sF}}{(2\,\pi)^{3}}\, \frac{{\rm d} {\rm {\bf p}}_{B}}{(2\,\pi)^{3}}\,\frac{{\rm d} {\rm {\bf p}}_{b}}{(2\,\pi)^{3}}                                               
% \frac{  {\overline \Phi}_{  (bB){\rm {\bf k}}_{B}\, {\rm {\bf k}}_{b} }^{(+)}\big({\rm {\bf p}}_{B},\,{\rm {\bf p}}_{b}\big)\,W_{bB}\big({\rm {\bf p}}_{bB} \big)\, \Psi_{ {\rm {\bf k}}_{sF} }^{C(-)}\big({\rm {\bf p}}_{sF}\big)\,{ M}_{\rm tr}\big({\rm {\bf k}}_{sF} ,\,{\rm {\bf k}}_{aA} ) }
%{k_{\rm  R}^{2} - k_{sF}^{2} },\\
%%
&{ M}'= - \frac{i\,\mu_{sF}}{2\,\pi} \,\int\,
%\frac{{\rm d} {\rm {\bf k}}_{sF}}{(2\,\pi)^{3}}\, 
\frac{{\rm d} {\rm {\bf p}}_{B}}{(2\,\pi)^{3}}\,\frac{{\rm d} {\rm {\bf p}}_{b}}{(2\,\pi)^{3}}                                               
 {  \Phi}_{{\rm {\bf k}}_{B}, {\rm {\bf k}}_{b} }^{(+)}\big({\rm {\bf p}}_{B},\,{\rm {\bf p}}_{b}\big)\,W_{bB}\big({\rm {\bf p}}_{bB} \big)\, 
 {  J}({\rm {\bf p}}_{sF},\, {\rm {\bf k}}_{aA}),
 %\frac{  
 %\Psi_{ {\rm {\bf k}}_{sF} }^{C(-)}\big({\rm {\bf p}}_{sF}\big)\,{ M}_{\rm tr}\big({\rm {\bf k}}_{sF} ,\,{\rm {\bf k}}_{aA} ) }
%{k_{\rm  R}^{2} - k_{sF}^{2} },
\label{THMres2}
\end{align} 
%\end{widetext}
where 
\begin{align}
{  J}({\rm {\bf p}}_{sF},\, {\rm {\bf k}}_{aA})   = 
\int \,
\frac{{\rm d} {\rm {\bf k}}_{sF}}{(2\,\pi)^{3}}\,
\frac{ \Psi_{ {\rm {\bf k}}_{sF} }^{C(-)}\big({\rm {\bf p}}_{sF}\big)\,{ M}_{\rm tr}\big({\rm {\bf k}}_{sF} ,\,{\rm {\bf k}}_{aA} \big) }
{  k_{\rm R(sF)}^{2}  - k_{sF}^{2} },
\label{J11int} 
\end{align}
and
\begin{align}
 { M}_{\rm tr} \big({\rm {\bf  k}}_{sF},\,{\rm {\bf k}}_{aA}\big)= \blb \Psi_{  {\rm {\bf k}}_{sF} }^{C(-)}\, {\tilde \phi}_{{\rm R}(xA)} \bl {\cal U}_{sA} \bl I_{x}^{a}\,\Psi_{  {\rm {\bf k}}_{aA}    }^{C(+)} \blk
 \label{M11}
 \end{align}
is the amplitude of the transfer reaction $a + A \to s + F^{*}$  populating the resonance $F^{*}$.
Also we introduced
\begin{align}
k_{{\rm R}(sF)}^{2}/(2\,\mu_{sF}) = E_{{\rm R}(sF)}= E_{0(sF)} + i\,\Gamma/2,
\label{k0sF1}
\end{align}
where 
\begin{align}
E_{0(sF)}= E_{aA} - \varepsilon_{a} + Q_{if}  - E_{0(bB)}.
\label{ER1}
\end{align}
Note that the imaginary parts  ${\rm Im} (k_{\rm R}) >0$  and  ${\rm Im} (E_{\rm R}) >0$.  
Then we have
\begin{align}
k_{{\rm R}(sF)}^{2}/(2\,\mu_{sF}) - k_{(sF)}^{2}/(2\,\mu_{sF}) = E_{0(sF)} - E_{(sF)} + i\,\Gamma/2
=E_{bB} - E_{0(bB)}  + i\,\Gamma/2.
\label{ERsFERbB1}
\end{align}

The expression for the form factor $W_{bB}\big({\rm {\bf p}}_{bB}\big)$ can be obtained 
 using Eq. (\ref{Gamowwfi1}) from  \cite{blokh84}:  
\begin{align}
W_{bB}\big({\rm{\bf p}}_{bB} )    &= \int {d{{\bf r}_{bB}}} {e^{-i{{\bf p}_{bB}} \cdot {{\bf r}_{bB}}}}{\widetilde V_{bB}}({{\bf r}_{bB}}){ \varphi _{{\rm R}(bB)}}({{\bf r}_{pB}})                            \nonumber\\
&={e^{i{\delta ^p}({k_{0(bB)}})}}\,{e^{\frac{{ - \pi {\eta _{{\rm R}(bB)}}}}{2}}}\,{\left[ {\frac{{p_{bB}^2 - k_{{\rm R}(bB)}^2}}{{4k_{{\rm R}(f)}^2}}} \right]^{i{\eta _{R(bB)}}}}\,
\Gamma (1 - i{\eta _{{\rm R}(bB)}})\,\sqrt{\frac{\mu_{bB}\,\Gamma_{bB}}{k_{{\rm R}(bB)}  }}\,g(p_{bB}^{2}),
 \label{formfactor1}
\end{align}
where $g(p_{bB}^{2})$   is the so-called reduced form factor, which satisfies $g(k_{{\rm R}(bB)}^{2})=1$.

We can single out now the resonance term of ${ M}'$.  First, we consider the integral \eqref{J11int}.
We represent the latter as
\begin{align}
{  J}({\rm {\bf p}}_{sF},\, {\rm {\bf k}}_{aA})   &=  4\,\pi\, \sum\limits_{lm_{l}}\, Y_{lm_{l}}^{*}({\rm {\bf {\hat p}}}_{sF})\,Y_{lm_{l}}({\rm {\bf {\hat k}}}_{aA})\,{  J}_{l}( p_{sF},\, k_{aA}).
\label{J11} 
\end{align}
In addition, we expand in partial waves 
%the amplitude of the transfer reaction and 
the Coulomb wave function as
%\begin{align}
%{ M}_{\rm tr}\big({\rm {\bf k}}_{sF} ,\,{\rm {\bf k}}_{aA} \big) =   4\,\pi\, \sum\limits_{lm_{l}}\, Y_{lm_{l}}^{*}({\rm {\bf {\hat k}}}_{sF})\,Y_{lm_{l}}({\rm {\bf {\hat k}}}_{aA})\,{ M}_{tr \, l}( k_{sF},\, k_{aA}),
%\label{Mtrl1}
%\end{align}
%and
\begin{align}
 \Psi_{ {\rm {\bf k}}_{sF} }^{C(-)}({\rm {\bf p}}_{sF} )\,=     \Psi_{- {\rm {\bf k}}_{sF} }^{C(+)*}({\rm {\bf p}}_{sF})   =
 \, 4\,\pi\, \sum\limits_{lm_{l}}\, Y_{lm_{l}}({\rm {\bf {\hat k}}}_{sF})\,Y_{lm_{l}}^{*}({\rm {\bf {\hat p}}}_{sF})\,\Psi^{C*}_{ k_{sF}, l}( p_{sF} ). 
 \label{Psi1}
\end{align}
%respectively. 
The partial-wave Coulomb scattering wave  function is given by \cite{dol1966}
\begin{align}
\Psi_{ k_{sF}, l}^{C*}( p_{sF} ) =& \frac{2 \pi i}{p_{sF}} e^{-\pi \eta_{k_{sF}}/2 } \Gamma(1-i \eta_{k_{sF}}) 
\nonumber \\ & \times 
\lim_{\beta \to +0} \Big[ (p_{sF} +  k_{sF}  -i \beta)^{-1-i \eta_{k_{sF}}}  (p_{sF} - k_{sF} -i \beta)^{ -1+ \eta_{k_{sF}}  }  \nonumber\\
& \times F\Big(-l, l+1; 1+i \eta_{k_{sF}}; -\frac{   (p_{sF}   -  k_{sF} )^{2} }{ 4 p_{sF} k_{sF}  }\Big)
\nonumber \\ & 
- e^{-2 i \alpha_{l}} (p_{sF} +  k_{sF} + i \beta)^{-1+i \eta_{k_{sF}}}  (p_{sF} - k_{sF} +i \beta)^{ -1- i \eta_{k_{sF}}  }      \nonumber\\
& \times  F \Big(-l, l+1; 1-i \eta_{k_{sF}}; -\frac{(p_{sF}   -  k_{sF} )^{2} }{4 p_{sF} k_{sF}} \Big) \Big],
 \label{Psil1}
\end{align}
where $F(a;b;c;z)$  is the hypergeometric function,  $ \alpha_{l}= \sigma_{l}^{C} -  \sigma_{0}^{C}$, and  $\sigma_{l}^{C}$
 is the Coulomb scattering wave function in the partial wave $l$.
 
We show now that the function ${  J}({\rm {\bf p}}_{sF},\, {\rm {\bf k}}_{aA}) $   has a resonance behavior when  $p_{sF}   \to  k_{ {\rm R}} $.  
The integrand  in Eq. (\ref{J11int})   has a pole at  $k_{sF} = k_{\rm R}$   in the $k_{sF}$ plane.  It is  located  in the upper half plane  and this is evident from Eq. (\ref{k0sF1}).  Besides, the integrand has four branching points  at $k_{sF} = \mp  p_{sF}   \pm  i\,\beta$.  When $p_{sF}  \to k_{\rm R}$
only the branching point $k_{sF}= p_{sF}  - i\,\beta$ moves from the fourth quadrant to the first one pinching the integration contour to  the pole  at $k_{sF}= k_{\rm R}$. 

Taking the residue at the pole $k_{sF} =k_{\rm R}$  one gets
\begin{align}
{  J}({\rm {\bf p}}_{sF},\, {\rm {\bf k}}_{aA}) = {  J}_{\rm R}({\rm {\bf p}}_{sF},\, {\rm {\bf k}}_{aA})  + \Delta\,{  J}({\rm {\bf p}}_{sF},\, {\rm {\bf k}}_{aA}), 
\label{JR1A}
\end{align}
where $\Delta\,{  J}({\rm {\bf p}}_{sF},\, {\rm {\bf k}}_{aA})$ is a nonresonant  term at $p_{sF} = k_{\rm R}$  and 
the resonant term is given as
\begin{align}
{  J}_{\rm R}({\rm {\bf p}}_{sF},\, {\rm {\bf k}}_{aA}) = e^{-\pi\,\eta_{\rm R}/2}\,\Gamma(1- i\,\eta_{{\rm R}})\,(2\,k_{\rm R})^{-2\,i\,\eta_{\rm R}}\,
{ M}_{\rm tr}\big( k_{\rm R}{\rm {\bf {\hat p}}}_{sF} ,\,{\rm {\bf k}}_{aA} \big)\,\big(p_{sF}^{2}  -  k_{\rm R}^{2}  \big)^{-1+ i\,\eta_{\rm R}},
\label{JR1}
\end{align}
$\eta_{\rm R}   = ( Z_{s}\,Z_{F}/137)\,\mu_{sF}/k_{\rm R}$ and   $\;{ M}_{\rm tr}\big( k_{\rm R}{\rm {\bf {\hat p}}}_{sF} ,\,{\rm {\bf k}}_{aA} \big)\,$
is given by  Eq. (\ref{M11})  in which  $\,{\rm {\bf k}}_{sF}$   is replaced by  $\,k_{\rm R}{\rm {\bf {\hat p}}}_{sF}$.
One very important observation is that, due to the Coulomb $s-F^{*}$ interaction in the intermediate state of the TH reaction described by the Coulomb scattering wave function  $\Psi_{ {\rm {\bf k}}_{sF} }^{C(-)}$, the resonance pole  at $\,p_{sF} = k_{\rm R}\,$  becomes a singular branching point.  

Replacing now in Eq. (\ref{THMres2}) $\,{  J}({\rm {\bf p}}_{sF},\, {\rm {\bf k}}_{aA})\,$  by  $\,{  J}_{\rm R}({\rm {\bf p}}_{sF},\, {\rm {\bf k}}_{aA})\,$  one gets
 \begin{align}                                     
{ M}_{\rm R}=& -e^{-\pi\,\eta_{\rm R}/2}\,\Gamma(1- i\,\eta_{{\rm R}})\,(2\,k_{\rm R})^{-2\,i\,\eta_{\rm R}}\, \frac{i\,\mu_{sF}}{2\,\pi} \,\int\, \frac{{\rm d} {\rm {\bf p}}_{B}}{(2\,\pi)^{3}}\,\frac{{\rm d} {\rm {\bf p}}_{b}}{(2\,\pi)^{3}}                                               
 \frac{  {  \Phi}_{{\rm {\bf k}}_{B}, {\rm {\bf k}}_{b} }^{(+)}\big({\rm {\bf p}}_{B},\,{\rm {\bf p}}_{b}\big)\,W_{bB}\big({\rm {\bf p}}_{bB} \big)\, 
  }
{\big(p_{sF}^{2}  -  k_{\rm R}^{2}  \big)^{1- i\,\eta_{\rm R}} }
\nonumber \\ & \times 
{\ M}_{\rm tr}\big(k_{\rm R}{\rm {\bf {\hat p}}}_{sF},\,{\rm {\bf k}}_{aA} \big).
\label{THMR1}
\end{align} 

Even before performing integrations over $\,{\rm {\bf p}}_{B}\,$   and $\,{\rm {\bf p}}_{b}\,$  one can figure out  how a resonance appears in function $\,{ M}_{\rm R}$.   Three-body Coulomb wave function  $\, {  \Phi}_{{\rm {\bf k}}_{B}, {\rm {\bf k}}_{b} }^{(+)}\big({\rm {\bf p}}_{B},\,{\rm {\bf p}}_{b}\big)\,$  has singularities  at $ \,{\rm {\bf p}}_{b} = {\rm {\bf k}}_{b}\,$ and $\,{\rm {\bf p}}_{B}= {\rm {\bf k}}_{B}$  corresponding to forward  scattering.
Taking into account that in  the  c.m. of the  TH reaction,  $\,{\rm {\bf p}}_{b} +  {\rm {\bf p}}_{B} + {\rm {\bf p}}_{s}  =0\,$ and $\, {\rm {\bf k}}_{b} + {\rm {\bf k}}_{B} + {\rm {\bf k}}_{s} =0\,$  we get that from $ \,{\rm {\bf p}}_{b} = {\rm {\bf k}}_{b}\,$ and $\,{\rm {\bf p}}_{B}= {\rm {\bf k}}_{B}\,$  follows $\,{\rm {\bf p}}_{s}= {\rm {\bf p}}_{sF}= {\rm {\bf k}}_{sF(f)}$, where $ {\rm {\bf k}}_{sF(f)}$ is the on-shell relative momentum of $s$ and the c.m. of the $b+B$ system in the final state.

Coincidence of the forward singularities of  $\, {  \Phi}_{{\rm {\bf k}}_{B}, {\rm {\bf k}}_{b} }^{(+)}\big({\rm {\bf p}}_{B},\,{\rm {\bf p}}_{b}\big)\,$  with the singularity at  $p_{sF} = k_{\rm R}$  leads  to  the resonant singularity of $\,{ M}_{\rm R}\,$  at 
$\,{\rm {\bf k}}_{s} = {\rm {\bf k}}_{sF} =  k_{\rm R}{\rm {\bf {\hat p}}}_{sF}$.  That is why, when looking for a resonance behavior of  $\,{ M}_{\rm R}\,$,   the amplitude  $\, { M}_{\rm tr}\big(k_{\rm R}{\rm {\bf {\hat p}}}_{sF},\,{\rm {\bf k}}_{aA} \big)\,$ can be taken out of the integral sign at  ${\rm {\bf  {\hat p}}}_{sF} = {\rm {\bf {\hat k}}}_{sF(f)}$.

To find an explicit expression for ${ M}_{\rm R}$  one does not need to know  the  three-body Coulomb scattering wave function 
$ {  \Phi}_{{\rm {\bf k}}_{B}, {\rm {\bf k}}_{b} }^{(+)}\big({\rm {\bf p}}_{B},\,{\rm {\bf p}}_{b}\big)$. To this end  the three-body Coulomb wave function can be replaced by the three-body CAS \cite{ashurov1984,muk1985}: 
\begin{align}
\blb P\big|K\infty \blk=&  \int\,\frac{{\rm d} {\rm {\bf p}}}{(2\,\pi)^{3}}\, \blb {\rm {\bf p}}_{b} - {\rm {\bf k}}_{b} - {\rm {\bf p}} 
+ {\rm {\bf k}}_{bB} +  {\rm {\bf k}}_{bs} \big| {\rm {\bf k}}_{bs}\,\infty \blk\,
\blb {\rm {\bf p}}_{B} - {\rm {\bf k}}_{B} + {\rm {\bf p}} - {\rm {\bf k}}_{bB} +  {\rm {\bf k}}_{Bs} \big| {\rm {\bf k}}_{Bs}\,\infty \blk\,
\nonumber \\ & \times 
\blb {\rm {\bf p}} \big|{\rm {\bf k}}_{bB}\,\infty \blk,
\label{ThreebodyCAS1}
\end{align}
where $\blb {\rm {\bf p}} \big| {\rm {\bf k}} \infty \blk$  is the two-body CAS \cite{vanHaer1976,muk1985}.  $P= \{ {\rm {\bf p}}_{bB},\,{\rm {\bf p}}_{sF} \}$
and $K= \{{\rm{\bf k}}_{bB},\,{\rm {\bf k}}_{sF(f)} \}$  are $6$-dimensional momenta,  ${\rm {\bf k}}_{ij} = (m_{j}\,{\rm {\bf k}}_{i} - m_{i}\,{\rm {\bf k}}_{j})/m_{ij}$
is the $i-j$ relative momentum,  ${\rm {\bf k}}_{i}$ and $E_{i} = k_{i}^{2}/(2\,m_{i})$  are  on-the-energy shell momentum and energy of  particle $i$ in the final state, ${\rm {\bf p}}_{i}$ is the virtual momentum of particle $i$ and ${\rm {\bf p}}_{ij}$ is the $i-j$ virtual relative momentum, ${\rm {\bf p}}$ is the integration variable,  $m_{ij}= m_{i} + m_{j}$. 

The two-body CAS  is  a generalized distribution, which has support only in one point 
${\rm {\bf p}} ={\rm {\bf k}} $, is defined on  a specific class of test functions \cite{vanHaer1976}.  The support of the three-body CAS in the $6$-dimensional momentum space is ${\rm {\bf p}}_{b} ={\rm {\bf k}}_{b}$  and  ${\rm {\bf p}}_{B} = {\rm {\bf k}}_{B}$.  Hence, when folding with the three-body CAS all  the regular functions at the support points can be taken out of the integral sign. 
Then at $k_{bB}^{2}  \to  k_{{\rm R}(bB)}^{2}$ we have
 \begin{align}                                     
{ M}_{\rm R}  \stackrel{k_{bB}^{2}  \to  k_{{\rm R}(bB)}^{2}}{=}& - {e^{i{\delta ^p}({k_{0(bB)}})}}\, e^{-\pi[\eta_{\rm R}  +   \eta_{{\rm R}(bB)}] /2}\,\Gamma (1 - i{\eta _{{\rm R}(bB)}})\,\Gamma(1- i\,\eta_{{\rm R}})\,(2\,k_{\rm R})^{-2\,i\,\eta_{\rm R}}\,\nonumber \\ & \times 
\sqrt{\frac{\mu_{bB}\,\Gamma_{(bB)}}{k_{{\rm R}(bB)}  }}                 
%\nonumber \\  & \times 
{ M}_{\rm tr}\big(k_{\rm R}{\rm {\bf {\hat k}}}_{sF(f)},\,{\rm {\bf k}}_{aA} \big)\, \frac{i\,\mu_{sF}}{2\,\pi}    
 \int\, \frac{{\rm d} {\rm {\bf p}}_{B}}{(2\,\pi)^{3}}\,\frac{{\rm d} {\rm {\bf p}}_{b}}{(2\,\pi)^{3}}                                               
 \frac{ \blb P \big| K\infty \blk }{\big(p_{sF}^{2}  -  k_{\rm R}^{2}  \big)^{1- i\,\eta_{\rm R}} } 
 \nonumber \\ & \times 
{\left[ {\frac{{p_{bB}^2 - k_{{\rm R}(bB)}^2}}{{4k_{{\rm R}(bB)}^2}}} \right]^{i{\eta _{{\rm R}(bB)}}}},
\label{THMR2}
\end{align} 
where  $\eta_{{\rm R}(bB)}= (Z_{b}\,Z_{B}/137)\mu_{bB}/k_{{\rm R}(bB)}$.

It is convenient first to integrate over ${\rm {\bf p}}_{b}$ and ${\rm {\bf p}}_{B}$ and then over ${\rm {\bf p}}$ [recall that the latter is required for calculating $\blb P\big|K\infty \blk$ entering Eq. \eqref{THMR2}, see Eq. \eqref{ThreebodyCAS1}].  Factor  $ {\left[ {({{p_{bB}^2 - k_{{\rm R}(bB)}^2}})/{{4k_{{\rm R}(bB)}^2}}} \right]^{i{\eta _{\rm R}}}} $  can be taken out  of the integrals at support points  of   two-body CASs
$ \blb {\rm {\bf p}}_{b} - {\rm {\bf k}}_{b} - {\rm {\bf p}} + {\rm {\bf k}}_{bB} +  {\rm {\bf k}}_{bs} \big| {\rm {\bf k}}_{bs}\,\infty \blk$ and $ \blb {\rm {\bf p}}_{B} - {\rm {\bf k}}_{B} + {\rm {\bf p}} - {\rm {\bf k}}_{bB} +  {\rm {\bf k}}_{Bs} \big| {\rm {\bf k}}_{Bs}\,\infty \blk$:
\begin{align}
{\rm {\bf p}}_{b} =  {\rm {\bf k}}_{b} + {\rm {\bf p}} - {\rm {\bf k}}_{bB},   \qquad
{\rm {\bf p}}_{B} =  {\rm {\bf k}}_{B} - {\rm {\bf p}} + {\rm {\bf k}}_{bB} .
\label{support2CAS1}
\end{align}
Then we have
\begin{align}
{\rm {\bf p}}_{bB} = \frac{m_{B}\,{\rm {\bf p}}_{b}  - m_{b}\,{\rm {\bf p}}_{B}}{m_{bB} }  =  {\rm {\bf p}}
\label{pbBp1}
\end{align}
and
\begin{align}                                     
{ M}_{\rm R}  \stackrel{k_{bB}^{2}  \to  k_{{\rm R}(bB)}^{2}}{=}&   - {e^{i{\delta ^p}({k_{0(bB)}})}}  e^{-\pi[\eta_{\rm R}  +   \eta_{{\rm R}(bB)}] /2} \Gamma (1 - i{\eta _{{\rm R}(bB)}}) \Gamma(1- i \eta_{{\rm R}}) (2 k_{\rm R})^{-2 i \eta_{\rm R}} 
\nonumber \\ & \times 
\sqrt{\frac{\mu_{bB} \Gamma_{(bB)}}{k_{{\rm R}(bB)}  }} 
%\\ & \times
{ M}_{\rm tr}\big(k_{\rm R}{\rm {\bf {\hat k}}}_{sF(f)}, {\rm {\bf k}}_{aA} \big)  \frac{i \mu_{sF}}{2 \pi} \int \frac{{\rm d} {\rm {\bf p}}}{(2 \pi)^{3}}  {\left[ {\frac{{p^2 - k_{{\rm R}(bB)}^2}}{{4k_{{\rm R}(bB)}^2}}} \right]^{i{\eta _{{\rm R}(bB)}}}}    
\nonumber\\ & \times  
\int  \frac{{\rm d} {\rm {\bf p}}_{B}}{(2 \pi)^{3}} \frac{{\rm d} {\rm {\bf p}}_{b}}{(2 \pi)^{3}}\frac{ \big<{\rm {\bf p}}_{b} - {\rm {\bf k}}_{b} - {\rm {\bf p}} 
+ {\rm {\bf k}}_{bB} +  {\rm {\bf k}}_{bs} \big| {\rm {\bf k}}_{bs} \infty \big>    }{\big[({\rm {\bf p}}_{b} +  {\rm {\bf p}}_{B})^{2}  -  k_{\rm R}^{2}  \big]^{1- i \eta_{\rm R}} } 
\nonumber \\ & \times 
\big<{\rm {\bf p}}_{B} - {\rm {\bf k}}_{B} + {\rm {\bf p}} - {\rm {\bf k}}_{bB} +  {\rm {\bf k}}_{Bs} \big| {\rm {\bf k}}_{Bs} \infty \big>    ,
\label{THMR3}
\end{align}
where in the c.m. of the TH reaction ${\rm {\bf p}}_{sF}  = {\rm {\bf p}}_{s}=  -{\rm {\bf p}}_{b} - {\rm {\bf p}}_{B}$.

Let us first consider the integral 
\begin{align}
{{  L}} = \int\, \frac{{\rm d} {\rm {\bf p}}_{B}}{(2\,\pi)^{3}}\,\frac{{\rm d} {\rm {\bf p}}_{b}}{(2\,\pi)^{3}}\frac{ \big<{\rm {\bf p}}_{b} - {\rm {\bf k}}_{b} - {\rm {\bf p}} 
+ {\rm {\bf k}}_{bB} +  {\rm {\bf k}}_{bs} \big| {\rm {\bf k}}_{bs}\,\infty \big>\, \big<{\rm {\bf p}}_{B} - {\rm {\bf k}}_{B} + {\rm {\bf p}} - {\rm {\bf k}}_{bB} +  {\rm {\bf k}}_{Bs} \big| {\rm {\bf k}}_{Bs}\,\infty \big>      }{\big[({\rm {\bf p}}_{b} +  {\rm {\bf p}}_{B})^{2}  -  k_{\rm R}^{2}  \big]^{1- i\,\eta_{\rm R}} } .
\label{L1}
\end{align}
To calculate this integral, as it has been shown in \cite{muk1985},  one needs to single out the most singular term of ${  L}$.  This is equivalent to the replacement of the two-body CAS by the corresponding Coulomb scattering wave functions. In other words, we consider the integral
\begin{align}
{{  L}_{1}} = \int {\frac{{d{{\rm {\bf p}}_b}}}{{{{(2\pi )}^3}}}}\, \int {\frac{{d{{\rm {\bf p}}_B}}}{{{{(2\pi )}^3}}}} \, \frac{{ \Psi_{{\rm {\bf k}}_{bs}}^{C(+)}({{\rm {\bf p}}_b} - {{\rm {\bf k}}_b}  + {{\rm {\bf k}}_{bs}}) }{ \Psi_{{\rm {\bf k}}_{Bs}}^{C(+)}({{\rm {\bf p}}_B} - {{\rm {\bf k}}_B}  + {{\rm {\bf k}}_{Bs}}) }}{{{{[{{({{\rm {\bf p}}_b} + {{\rm {\bf p}}_B})}^2} - {\rm {\bf k}}_{\rm R}^2]}^{  1 - i{\eta _{\rm R}}}}}}.
\label{L2A}
\end{align}
Here, in the arguments of the Coulomb scattering wave functions we assumed ${\rm {\bf p}}= {\rm {\bf k}}_{bB}$  because it is support point of  the CAS  $\big<{\rm {\bf p}} \big| {\rm {\bf k}}_{bB}\big>$.

\begin{figure}[htbp]
\includegraphics[width=0.5\textwidth]{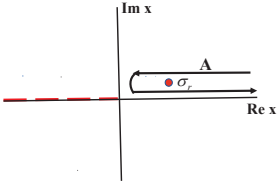}
\caption{Integration contour in the $x$-plane in Eq. (\ref{inttransf1}). The red point is a pole of the integrand at $x= \sigma_{r}$.
The red dotted line is the cut of the function $x^{-1 +i\,\eta_{\rm R} }$ going from $x=0$ to $-\infty$. The integration contour $A$
is shown by the thick solid line chosen so that there are no singularities inside this contour except for the pole at $x= \sigma_{r}$.  }
\label{fig_intcontour1}
\end{figure}

To proceed further one can use the Cauchy's theorem. To this end we introduce  the integral  
\begin{align}
\frac{1}{  \sigma_{r}^{ 1 - i\,\eta _{\rm R}  }}= - \frac{1}{2\,\pi\,i}\,\oint\limits_{A}\,{\rm d}x\,x^{-1 +i\,\eta_{\rm R} }\,\frac{1}{  (\sigma_{r}  -x )},
\label{inttransf1}
\end{align}
which is taken along a closed contour $A$. The contour begins at $x=\infty$, encircles the pole of the integrand  $\,x= \sigma_{{r}}$, where 
\begin{align}
\,\sigma_{r} = ({\rm {\bf p}}_b + {\rm {\bf p}}_B)^2 - {\rm {\bf k}}_{\rm R}^2,  
\label{sigmar1}
\end{align}
and goes back to $\,x=\infty$, see Fig. \ref{fig_intcontour1}.  
We can select the cut of the function $x^{-1 +i\,\eta_{\rm R} }$ going from $x=0$ to $-\infty$. Then there is only a pole singularity at $x=\sigma_{r}$ inside the integration contour $A$.  
 
Then we have
\begin{align}
{{  L}}_{1} = -\frac{1}{2\,\pi\,i}\,\int {\frac{{d{{\rm {\bf p}}_b}}}{{{{(2\pi )}^3}}}}\, \int {\frac{{d{{\rm {\bf p}}_B}}}{{{{(2\pi )}^3}}}} \,\oint\limits_{A}\,{\rm d}x\,x^{-1 +i\,\eta_{\rm R} }\,   \frac{{ \Psi_{{\rm {\bf k}}_{bs}}^{C(+)}({{\rm {\bf p}}_b} - {{\rm {\bf k}}_b}  + {{\rm {\bf k}}_{bs}}) }{ \Psi_{{\rm {\bf k}}_{Bs}}^{C(+)}({{\rm {\bf p}}_B} - {{\rm {\bf k}}_B}  + {{\rm {\bf k}}_{Bs}}) }}
{{({\rm {\bf p}}_b + {\rm {\bf p}}_B)}^2 - {\rm {\bf k}}_{\rm R}^2 - x}. 
\label{L2}
\end{align}
To calculate  this integral  we need to rewrite the integrand in the coordinate space:
\begin{align}
 {  L}_{1} =& - e^{{ -\pi(\eta_{Bs} + \eta_{bs} )/2 }} \,\Gamma(1 + i\,\eta_{Bs})\,\Gamma(1+ i\,\eta_{bs})\, \frac{1}{2\,\pi\,i}\, \int {\frac{{d{{\rm {\bf p}}_b}}}{{{{(2\pi )}^3}}}}\, \int {\frac{{d{{\rm {\bf p}}_B}}}{{{{(2\pi )}^3}}}} \,\oint\limits_{A}\,{\rm d}x\,x^{-1 +i\,\eta_{\rm R} }\, 
 \nonumber \\ & \times 
\int{\rm d}{\rm {\bf r}}\,e^{-i\,{\rm {\bf k}}_{s} \cdot {\rm {\bf r}}} \, \frac{e^{-\kappa\,r} }{r}                                                              %\nonumber\\ &  \times 
{}_1{F}_{1}\big(-i{\eta _{Bs}},1;i(k_{Bs}r - {\rm {\bf k}}_{Bs} \cdot {\rm {\bf r}}) \big)\,{}_1{F}_{1}\big(-i{\eta _{bs}},1;i({k_{bs}r - {\rm {\bf k}}_{bs}} \cdot {\rm {\bf r}}) \big).
\label{LC1}
\end{align}
Here we took into account  that the Coulomb scattering wave function in the coordinate space is given by 
\begin{align}
\Psi_{{\rm {\bf k}}}^{C(+)}({\rm {\bf r}}) =
e^{{ -\pi\,\eta/2 }} \,\Gamma(1 + i\,\eta)\,{}_1{F}_{1}\big(-i{\eta},1;i(k\,r - {\rm {\bf k}} \cdot {\rm {\bf r}}) \big),
\end{align}
 and utilized a Fourier transformation
\begin{align}
\frac{1}{{{\sigma _r} - x}} = \frac{1}{({\rm {\bf p}}_{b} + {\rm {\bf p}}_{B})^{2}  - k_{{\rm R}}^{2}  -x} =  \frac{1}{4\,\pi}\,\int \, {\rm{d}}{\bf{r}}\,e^{  i\,({\rm {\bf p}}_{b} + {\rm {\bf p}}_{B}) \cdot {{\rm {\bf r}}   }}\,\frac{{{e^{ - \kappa \,r}}}}{r}.
\label{Fouriersigmar1}
\end{align}
where $\kappa^{2}= \,- k_{{\rm R}}^{2}  -x$.  Note also that  in the c.m. of the TH reaction ${\rm {\bf p}}_{s}= -{\rm {\bf p}}_{b}- {\rm {\bf p}}_{B}$  and ${\rm {\bf p}}_{s}= {\rm {\bf p}}_{sF}$.

Despite the fact that $\sigma_{r}$  depends on the  ${\rm {\bf p}}_{b} + {\rm {\bf p}}_{B}$, the integration contour $A$  in Fig. \ref{fig_intcontour1} can be chosen so that it does not depend  on the integration variables ${\rm {\bf p}}_{b}$ and ${\rm {\bf p}}_{B}$. 
That is why we can change the integration order in Eq. (\ref{LC1}) and first integrate over momentum variables and then over ${\rm {\bf r}}$.    
 Using  the Nordsieck integral \cite{nordsieck}   one gets 
 %for ${  L}_{1}$  
\begin{align}
{  L}_{1} =& -\frac{1}{4\,\pi\,i}e^{{ -\pi(\eta_{Bs} + \eta_{bs} )/2 }} \,\Gamma(1 + i\,\eta_{Bs})\,\Gamma(1+ i\,\eta_{bs})\,\oint\limits_{A}\,{\rm d}x\,x^{-1 +i\,\eta_{\rm R} } \frac{{1 }}{\alpha }\,{\left[ {\frac{\alpha }{\gamma(\alpha) }} \right]^{-i{\eta _{bs}}}}\nonumber\\
&\times 
{\left[ {\frac{{\alpha  + \nu}}{\alpha }} \right]^{ i{\eta _{Bs}}}}
F\Big( - i{\eta _{Bs}},\,-i{\eta _{bs}},1;\frac{{ \alpha\, \tau -\nu\,\gamma( \alpha) }}{{\gamma(\alpha) (\alpha  + \nu)}} \Big),
\label{L11}
\end{align}
where $\alpha= 1/2\big(\sigma_{r} -x \big),$   $\;\nu= - {\rm {\bf k}}_{Bs} \cdot  {\rm {\bf  k}}_{s}  - i \,\kappa\, k_{Bs},$
$\gamma(\alpha)= -{\rm {\bf k}}_{bs} \cdot  {\rm {\bf k}}_{s} - i\,\kappa\,k_{bs} + \alpha\,$   and  $\;\tau  = k_{bs}\,k_{Bs}  + {\rm {\bf k}}_{bs} \cdot  {\rm {\bf k}}_{Bs} - \nu$.  Note that $\;\sigma_{r}$  is now given by 
\begin{align}
\sigma_{r}= k_{sF(f)}^{2} -  k_{{\rm R}}^{2}. 
\label{newsigmar1}
\end{align}

We can find the behavior of ${  L}_{1}$ at $\sigma_{r} \to 0\,$  what corresponds to approaching the resonance in the subsystem $F$. The singular behavior of the integral over $x$  at $\sigma_{r} \to 0$  is determined by the factor  $\alpha^{-1- \eta_{bs} - \eta_{Bs}}$. 
Using the substitution  $\,x=\sigma_{r} \,y\,\,$ we can rewrite Eq. (\ref{L11})  as
\begin{align}
{  L}_{1} \stackrel{\sigma_{r} \to 0}{=}&  - \frac{1}{4\,\pi\,i}e^{{ -\pi(\eta_{Bs} + \eta_{bs} )/2 }} \,\Gamma(1 + i\,\eta_{Bs})\,\Gamma(1+ i\,\eta_{bs})\,
\frac{1}{{\sigma _r^{1 + i[{\eta _{bs}} + {\eta _{Bs}} - {\eta _R}]}}}\, [\gamma(0)]^{i\,\eta_{bs}}\,\nu^{i\,\eta_{Bs}}\,   \nonumber\\
& \times F\big( - i{\eta _{Bs}},\,-i{\eta _{bs}},1;\,-1 \big)\, \oint\limits_{1}^{\infty}\,{\rm d}y\,y^{-1 +i\,\eta_{\rm R} }\,\frac{1}{(1- y)^{1 + i[\eta_{bs} + \eta_{Bs}]}}  .
\label{Lsigmar01}
\end{align}
Here, in all the factors regular at $\,\sigma_{r} \to 0\,$   we took $\,\sigma_{r}=0$.  

Integral over $y$ can be calculated analytically.  It is assumed that ${\rm arg}(y-1) =0$ at $y>1$. Then ${\rm arg}(1-y)= -\pi$ on the upper part of the contour $A$  going from $\,y=\infty\,$ to $y=1$  and  $\,{\rm arg}(1-y)= \pi\,$ for the lower part of the contour $A$ going from $\,y=1\,$  to $\,y=\infty\,$. 
Using Eq. (3.196(2)) from \cite{GR1} one gets 
\begin{align}
{  J}_{1}&=  -\oint\limits_{1}^{\infty}\,{\rm d}y\,y^{-1 +i\,\eta_{\rm R} } \frac{1}{(1- y)^{1 + i[\eta_{bs} + \eta_{Bs}]}}              \nonumber\\
&=-2\,{\rm sh}(\pi\,[\eta_{bs}+\eta_{Bs}]) \,\int\limits_{1}^{\infty}\, {\rm d}y\,y^{-1 +i\,\eta_{\rm R} } \frac{1}{(y-1)^{1 + i[\eta_{bs} + \eta_{Bs}]}}     \nonumber\\
&=-2\,{\rm sh}(\pi\,[\eta_{bs}+\eta_{Bs}])\,\frac{\Gamma\big(1+i[\eta_{bs} + \eta_{Bs}  -  \eta_{{\rm R}}] \big)\,\Gamma \big(-i[\eta_{bs} + \eta_{Bs}] \big)}{\Gamma \big(1- i\,\eta_{{\rm R}} \big)   }        
\label{oint1}
\end{align}
Taking into account the fact that
\begin{align}
{\rm sh}(-\pi\,\eta) = -\frac{i\,\pi}{\Gamma(1+i\,\eta)\,\Gamma(-i\,\eta)}
\label{shG1}
\end{align}
and  $\,{\rm arg}[\gamma(0)] =-\pi\,$  and $\,{\rm arg}(\nu) = -\pi$  we can rewrite ${  L}_{1}$ as
\begin{align}
{  L}_{1} \stackrel{\sigma_{r} \to 0}{=}& \frac{1}{2}e^{{ \pi(\eta_{Bs} + \eta_{bs} )/2 }} \,\Gamma(1 + i\,\eta_{Bs})\,\Gamma(1+ i\,\eta_{bs})\,                                                       \nonumber\\
& \times \frac{\Gamma\big(1+i[\eta_{bs} + \eta_{Bs}  -  \eta_{{\rm R}}] \big)}
 {\Gamma \big(1- i\,\eta_{{\rm R}} \big)\,\Gamma\big(1+i[\eta_{bs} + \eta_{Bs}] \big)   }  \frac{1}{{\sigma_r^{1 + i[{\eta _{bs}} + {\eta _{Bs}} - {\eta _R}]}}}\, [-\gamma(0)]^{i\,\eta_{bs}}\,(-\nu)^{i\,\eta_{Bs}}\,   \nonumber\\
&  \times F\big( - i{\eta _{Bs}},\,-i{\eta _{bs}},1;\,-1 \big).
\label{Lsigmarm3}
\end{align}

The remaining integral to be calculated is 
\begin{align}
W_{bB}\big({\rm{\bf k}}_{{\rm R}(bB)   } \big)  =& \sqrt{\frac{\mu_{bB}\,\Gamma_{(bB)}}{k_{{\rm R}(bB)}  }}\, 
e^{i\delta^{p}(k_{0(bB)}}\,e^{-\pi\,\eta_{{\rm R}(bB)}}\,\Gamma(1-i\,\eta_{{\rm R}(bB)}) \int\,\frac{{\rm d} {\rm {\bf p}}}{(2\,\pi)^{3}}\, \nonumber\\
& \times {\left[ {\frac{{p^2 - k_{{\rm R}(bB)}^2}}{{4k_{{\rm R}(bB)}^2}}} \right]^{i{\eta _{{\rm R}(bB)}}}}\, \big<{\rm {\bf p}} \big| {\rm {\bf k}}_{{\rm R}bB} \big>.
\label{JbB1}
\end{align}
The vertex form factor  $W_{bB}\big({\rm{\bf p}} \big)$, containing the Coulomb interaction, does not have an on-the-energy-shell limit at $p \to k_{{\rm R}(bB)}$    due to  the presence  of the factor  ${\left[ {{{(p^2 - k_{{\rm R}(bB)}^2})}/{{4k_{{\rm R}(bB)}^2}}} \right]^{i{\eta _{{\rm R}(bB)}}}}$. The two-body CAS plays a role of the renormalization factor. Folding of the vertex form factor with the CAS provides  on-the-energy-shell form factor
$W_{bB}\big({\rm{\bf k}}_{{\rm R}(bB) } \big) $.   To show it we can use  Eq. (16) from \cite{vanHaer1976}. Extension of this equation for the resonance momentum ${\rm {\bf k}}_{{\rm R}(bB)}$ gives:
%\begin{widetext}
\begin{align}
 \big< {\rm {\bf p}} \big|{ {\rm {\bf k}}_{{\rm R}(bB)}} \big>  = \frac{1}{{{e^{ - \pi \,{\eta _{{\rm R}(bB)}}}}\,\Gamma (1 - i\,{\eta _{\rm R}(bB)}){{(2{k_{{\rm R}(bB)}})}^{ - i{\eta _{{\rm R}(bB)}}}}}}\,           
 \frac{\delta({\rm {\bf p}}  - {\rm {\bf k}}_{{\rm R}bB})}{{{{(p - {k_{{\rm R}(bB)}})}^{i\,{\eta _{{\rm R}(bB)}}}}}}.
\label{CASR1}
\end{align}
%\end{widetext}
%
Substituting Eq. (\ref{CASR1}) into Eq. (\ref{JbB1})  we get
\begin{align}
W_{bB}\big({\rm{\bf k}}_{{\rm R}(bB)   } \big) = e^{i\,\delta^{p}(k_{0(bB)})} \sqrt{\frac{\mu_{bB}\,\Gamma_{(bB)}}{k_{{\rm R}(bB)}  }}.
\label{WbBR1}
\end{align}

Collecting all the factors we obtain the final equation for the THM reaction amplitude 
\begin{align}                                     
{ M}_{\rm R}  \stackrel{k_{bB}^{2}  \to  k_{{\rm R}(bB)}^{2}}{=}& - \,\frac{ \Gamma(1+ i\,\eta_{bs})\,\Gamma(1 + i\,\eta_{Bs})}
 {\Gamma\big(1+i[\eta_{bs} + \eta_{Bs}] \big)   }\,F\big( - i{\eta _{Bs}},\,-i{\eta _{bs}},1;\,-1 \big) 
 [-\gamma(0)]^{i\,\eta_{bs}}\,(-\nu)^{i\,\eta_{Bs}}\,   \nonumber\\
&  \times \,\sqrt{\frac{\mu_{bB}\,\Gamma_{(bB)}}{k_{{\rm R}(bB)}  }}\,\frac{i\,(2\mu_{sF})^{- i\,\zeta} }{4\,\pi} \,\frac{\Gamma\big(1+i\zeta \big)\,e^{{ \pi\,\zeta/2 }} \, }{{{(E_{{\rm R}(bB)}  - E_{bB} })^{1 + i\,\zeta}}}\,{ M}_{\rm tr}\big(k_{\rm R}{\rm {\bf {\hat k}}}_{sF},\,{\rm {\bf k}}_{aA} \big),
\label{THMRfinal1}
\end{align}
where
\begin{align}
\zeta=\eta _{bs} + \eta _{Bs} - \eta _{\rm R},
\label{zeta1}
\end{align}
$E_{sF(f)}  -   E_{{\rm R}}= E_{{\rm R}(bB)} -  E_{bB(f)} = E_{0(bB)} - E_{bB(f)} - i\,\Gamma/2 $,    
$\, E_{{\rm R}}= k_{{\rm R}}^{2}/(2\,\mu_{sF})$, $k_{{\rm R}}$ is given by Eq. (\ref{k0sF1}),
$E_{sF(f)}$ is  relative kinetic energy of the particle $s$ and the c.m. of the system $b+B$ in the final state, and
$E_{bB(f)}$ is the $b-B$ relative kinetic energy in the final state.

\section{Discussion}

Using a few-body formalism we have derived an expression for the amplitude of the TH reaction $a + A \to s+ F^{*} \to s + b +B$ proceeding though a resonance in the intermediate binary subsystem $s+ F^{*}$.
The Coulomb interactions in the intermediate binary and final three-body states have been taken into account explicitly using the three-body approach.  The following important conclusions can be drawn:

\begin{enumerate}
\item
The amplitude  of the TH reaction proceeding through the intermediate resonance $F$ in the binary subsystem 
has in the denominator the resonant energy factor  $E_{0(bB)} - E_{bB(f)} -i\,\Gamma/2$. However, a conventional Breit-Wigner resonance pole 
$[ E_{0(bB)} - E_{bB(f)} - i\,\Gamma/2]^{-1}$ is converted into the branching point singularity
 $[E_{0(bB)} - E_{bB(f)} -i\,\Gamma/2]^{-1- i\,\zeta}$. This transformation of the resonance behavior of the TH reaction amplitude 
 is caused by the Coulomb interaction of the particle $s$ with the resonance in the intermediate state and with products $b$ and $ B$ of the resonance.

\item 
The reaction amplitude can be rewritten as   
\begin{align}
{ M}_{\rm R}=& -\frac{ i}{4\,\pi}\,\sqrt{\frac{\mu_{bB}}{k_{{\rm R}(bB)}  }}
\frac{e^{i\,\delta^{p}(k_{0(bB)})}\Gamma_{bB}^{1/2}\,{ M}_{\rm tr} }{E_{0(bB)} -E_{bB(f)} - i \frac{\Gamma}{2}}\,N_{C}(E_{bB(f)},\,\zeta),
\label{RAL1}
\end{align}
where
\begin{align}
N_{C}  =& \frac{ \Gamma(1+ i\,\eta_{bs})\,\Gamma(1 + i\,\eta_{Bs})}
 {\Gamma\big(1+i[\eta_{bs} + \eta_{Bs}] \big)   }\,F\big( - i{\eta _{Bs}},\,-i{\eta _{bs}},1;\,-1 \big) 
 [-\gamma(0)]^{i\,\eta_{bs}}\,(-\nu)^{i\,\eta_{Bs}}
 \nonumber\\
& \times \left[E_{0(bB)} -E_{bB(f)} - i \frac{\Gamma}{2}\right]^{-i\,\zeta}
 \label{N1}
\end{align}
is the Coulomb renormalization factor which is equal to unity when the Coulomb interactions are turned off.    
  
\item 
As we have mentioned in Introduction, the 
%three 
final-state Coulomb effects have an universal feature and should be taken into account 
whenever one considers nuclear or atomic reactions leading to the three-body final states. 
 If $\,\eta_{bs},\,\eta_{Bs}\,$ and $\,{\rm Re}(\eta_{{\rm R}})\,$ have the same sign, the Coulomb $\,s-F^{*}\,$ interaction in the intermediate state weakens the impact of the final state Coulomb $s-b$ and $s-B$ interactions  because the intermediate state Coulomb parameter $\eta_{\rm R}$ is subtracted from the final-state Coulomb parameters $\,\eta_{bs}+\eta_{Bs}$, see Eq. (\ref{zeta1}).
For example, if the Coulomb $s-F^{*}$ interaction in the intermediate state is turned off, that is $\eta_{{\rm R}} =0$, 
the resonance behavior of the TH reaction amplitude coincides with that from the papers \cite{Senashenko,Godunov}, where  the angular and energy  dependences  of the electrons ejected from autoionizing resonances induced by collisions of fast protons with atoms were investigated.

\item
The triply differential cross section for the TH reaction is given by
\begin{align}
\frac{{{d^3}\sigma }}{{d{\Omega _{{{\bf k}_{sF(f)}}}}d{\Omega _{{{\bf k}_{bB(f)}}}}d{E_{sF(f)}}}} = {\sigma _0}\frac{{{\Gamma _{bB}}}}{{[E_{0(bB)} - E_{bB(f)}]^{2} + {{{\Gamma ^2}}}/{4}}} \, \frac{{\pi \zeta }}{{{\rm sh}\zeta }}
\,{e^{ 2\zeta \arctan \frac{{2{\cal E}}}{\Gamma }}},
\label{TripleDCS1}
\end{align}
where $\,\Omega _{{{\bf k}_{ij(f)}}}\,$ is the solid angle corresponding to the direction of the relative momentum ${\rm {\bf k}}_{ij(f)}$
of the partiles $i$ and $j$ in the final state, $\, {\cal E}= E_{0(bB)} - E_{bB(f)}= E_{sF(f)}  - E_{0{(sF(f))}} $. 
Here we singled out the part of the triply differential cross section that determines the resonance line shape, width, shift and peak value 
as functions of the Coulomb factors. This part appears due to the Coulomb interactions in the intermediate and final states.
The factor $\sigma_{0}$  does not affect the resonant line shape. 
The parameter $\zeta$ plays a crucial role in the modification of the resonance line shape. In particular, the shift of the resonance peak caused by the Coulomb interaction  is given by
\begin{align}
\Delta E_{0(bB)} = \frac{1}{2}\,\zeta\,\Gamma.
\label{deltaE1}
\end{align}

At large $\zeta$ the resonance peak decreases as $|\zeta|^{-1}$. 
The peak value of the resonant triply differential cross section is given as
\begin{align}
\frac{{{d^3}\sigma }}{{d{\Omega _{{{\bf k}_{sF(f)}}}}d{\Omega _{{{\bf k}_{bB(f)}}}}d{E_{sF(f)}}}} = {\sigma _0}\frac{{4{\Gamma _{bB}}}}{{{\Gamma ^2}(1 + {\varsigma ^2})}} \, \frac{{\pi \zeta }}{{{\rm sh}\zeta }} \, {e^{2\zeta \arctan \varsigma }}.
\label{peaktriplediffrsect1}
\end{align}

\item
If one of the Coulomb parameters in the final state, for example $\eta_{bs}$, is zero or very small then the cumulative effect of the Coulomb interactions in the intermediate and final state is negligible. 

\item
Our Eq. (\ref{TripleDCS1}) coincides with the one obtained in Ref. \cite{Kuchiev}   where atomic resonant processes were investigated
in the eikonal approach. However, here we obtained this equation using a general approach keeping in mind application for the indirect THM in nuclear physics.  Coincidence of our result and those given in Ref. \cite{Kuchiev} demonstrates a universality of the three-body Coulomb final-state effects in the resonant reactions. 

\end{enumerate}

Concluding, in this work we discussed the effect of the Coulomb interactions in the intermediate and final states on the resonant line shape. For the application of the THM it is also important to know the energy dependence of the triply differential cross section. In addition, it requires calculations of the energy dependence of the transfer cross section, which can play a crucial role on the energy.  These will be considered elsewhere.

\section*{Acknowledgements}
A.S.K. acknowledges a support from the Australian Research
Council. A.M.M. acknowledges a support from the U.S. DOE
Grant No. DE-FG02-93ER40773, the U.S. NSF Grant No.
PHY-1415656, and the NNSA Grant No. DE-NA0003841.

\end{document}